\documentstyle[prl,eqsecnum,aps]{revtex}

\begin{document}

\title{Quantum Relaxation of Magnetisation in Magnetic Particles
  }

\author{N. V. Prokof'ev$^{1,2}$  and P. C. E. Stamp$^{2}$}
\address{
$^{1}$ Russian Science Center "Kurchatov Institute", Moscow 123182, Russia\\
$\;\;\;$ \\
$^{2}$ Physics Department, University of British Columbia, 6224 Agricultural
Rd.,\\
 Vancouver B.C., Canada V6T 1Z1 \\ }
\maketitle

\vspace{1cm}
\begin{abstract}
At temperatures below the magnetic anisotropy energy,
monodomain magnetic systems
(small particles, nanomagnetic devices, etc.) must relax quantum
mechanically - thermal activation is ineffective. The discrete nature  of the
spectrum is important. This quantum relaxation must be mediated by the coupling
to both nuclear spins and phonons (and electrons if either particle or
substrate is conducting).

We analyze the effect of each of these couplings, and then combine
them. Conducting systems can be modelled  by a "giant Kondo" Hamiltonian,
with nuclear spins added in as well. At low temperatures, even microscopic
particles on a conducting substrate (containing only $10-50$ spins) will
have their magnetisation frozen over millenia by a combination of electronic
dissipation and the "degeneracy blocking" caused by nuclear spins. Raising
the temperature leads to a sudden unblocking of the spin dynamics at a well
defined temperature.

Insulating systems are quite different. The relaxation is strongly
enhanced by the coupling to nuclear spins. At short times the magnetisation
of an ensemble of particles relaxes logarithmically in time, after
an initial very fast decay - this relaxation proceeds entirely via the nuclear
spins. At longer times phonons take over, but the decay rate is still
governed by the temperature-dependent
nuclear bias field acting on the particles - decay may be exponential
or power-law depending on the temperature.

The most surprising feature of the results is the pivotal role played
by the nuclear spins. The results are relevant to any experiments on
magnetic particles in which interparticle dipolar interactions are unimportant.
They are also relevant to future magnetic device technology.

\end{abstract}

\vspace{3mm}

\section{Introduction}
\label{sec:in}

\subsection{Quantum Relaxation}

One of the most thoroughly explored subjects in all of science is that
of thermal relaxation of magnetisation, in magnetic systems of all
shapes and sizes.
Investigations in this area (which go back many centuries \cite{1})
have revealed many subtleties, and even today there are many unsolved
puzzles (e.g., the physics of "magnetic avalanches").

Very recently a whole new set of questions in this area has arisen,
with the advent of well-characterised "nanoscopic" magnetic structures
\cite{2}.
Such structures include "made-to-order" magnetic grains, magnetic wires and
superlattice arrays, as well as thin films and spin chains. There are very
obvious applications of such nanomagnets in, e.g., the recording industry,
as well as in information transmission and computing. Theoretical activity
has been particularly intense on 1-dimensional
spin systems \cite{3}, and on the tunneling and coherence phenomenon which
should exist in both grains (involving "giant spin" \cite{4,5} dynamics)
and in
the dynamics of domain walls \cite{6}. Both the theory and the experimental
activity in the latter field have been reviewed recently \cite{7,8,9}.

In this paper we focus on something rather different -
the "quantum relaxation" of magnetisation that
must occur in any small magnetic particle, once thermally-activated
relaxation has ceased. Although there does not seem to
have been a serious theoretical analysis of such relaxation before, the
  conventional view has been that
it must proceed via a simple tunneling process (dissipative or otherwise).
The picture we shall derive here is
 more subtle, and contains some rather fascinating new physics.
 Tunneling is certainly involved; but we find that
 what really controls relaxation is the nuclear spin system inside the
magnetic particle, often with the help of phonons or electrons. One can
put the result in the following way:
the nuclear spins provide the essential "switch" that turns on
(or turns off) the quantum
transitions. As far as we are aware, there has been no previous recognition
that nuclei had any role whatsoever to play in magnetic relaxation, despite
the enormous number of experiments in this area.
Nor do we know of any theoretical analysis of the influence
of electrons on magnetic relaxation. There has certainly been no
attempt at a theory dealing with the combined effects of nuclear spins,
phonons, and electrons, which is what we do here. As we shall see,
it is impossible to give a realistic analysis without putting them all
together.

We shall not try here to address experiments in any detail, for two reasons.
First, each different experimental system has its own idiosyncracies,
and we shall see that our theory has many different parameters - it is
simply not possible to give detailed results for all possible cases, and we
feel that it would be preferable to examine particular systems for which
parameters are well known, as test cases. This we intend to do elsewhere.
Second, in most experiments on magnetic grains, the grains are rather
close together. As we shall briefly explain in this paper, this makes it
probable that even the low-T relaxation of an experimental ensemble of grains
proceeds via dipolar interactions between the grains.

The theory in this paper is closely related to previous work on
quantum coherence in magnetic grains \cite{10}. In that work a very detailed
analysis was given of the way nuclear spins control the dynamics of
"giant spins", i.e., spins with quantum number $S \gg 1$. We also analysed
the effect of phonons and electrons in "decohering" the quantum motion
of $\vec{S}$. The problem of quantum relaxation is  different, in that
even if coherence is destroyed, relaxation may still proceed incoherently.
What we shall find is that in the absence of nuclear spins, phonons cannot
relax $\vec{S}$ in a small biasing field, because they couple too weakly to it;
on the other hand electrons couple so strongly to $\vec{S}$ that they
freeze it completely. The nuclear spins can then {\it liberate} $\vec{S}$ -
amusingly, the mechanisms by which they do this are the same as those by which
they {\it destroy} the quantum coherent motion of $\vec{S}$. The bias field
$\epsilon$ acting on $\vec{S}$, generated by the combined hyperfine fields of
all the nuclei can easily be greater than   $10 \:K$
(e.g., for TbFe$_3$ grains), and
depending on the size of the grain may be equivalent to an external field from
$100 \:G$ up to well over $1\: T$! This bias changes with time in
a diffusive way, at a rate governed by inter-nuclear dipolar interactions
(in the absence of phonons and electrons), and also
nuclear spin-phonon or spin-electron interactions, when
the latter are present. Every so often, as $\epsilon $ shunts around, it brings
quantum 2 states corresponding to different orientations of $\vec{S}$ into near
degeneracy. At this point the system can tunnel, provided (a) the nuclear bias
field $\epsilon (t)$ gives it time to do so coherently, (b) the usual
dissipative effects of the electrons and phonons are not too destructive.

On the other hand, we shall also see that the nuclear spins can also
further {\it inhibit} the relaxation of $\vec{S}$, at low $T$. If the
nuclear spins are allowed to come to equilibrium, then at temperature
below the hyperfine coupling energy, they exert a large negative bias field
on $\vec{S}$, which essentially traps $\vec{S}$ for astronomical times.
Thus, depending on $T$ and on the nuclear polarisation, the nuclei
can enhance or suppress relaxation.

Although our purpose in this paper is not to deal with the implications
for magnetic technology involving quantum relaxation, we feel we should
at least make some remarks here about what they might be. Fig.1 shows
the way in which magnetic computer memory elements have decreased
exponentially in size in the last 40 years - such plots have figured large
in recent discussions of the need for new kinds of computer design,
incorporating elements which operate quantum-mechanically. At
present computers, as well as magnetic tapes, use elements which behave
classically, and are stable over long periods of time. This
stability over decades exists because they are {\it big} - the energy barrier
between 2 states of the element, proportional to its size, usually
exceeds $100\: k_BT$ even at room temperature. The hypothesized
"quantum threshold", below which tunneling is important even at $T=0$,
is usually supposed to be for grains containing roughly $10^2-10^4$ spins,
depending on the material involved. However we shall find that the threshold
between stable behaviour and quantum relaxation can be moved around
a lot, depending on how the magnetic system is coupled to the nuclear spins,
and whether or not  it is coupled to electrons. We shall see that in some
cases, one can freeze the dynamics of even microscopic magnetic systems
(containing only 10 spins), for times of thousands of years or more.
Conversely, we shall see how it is possible for much larger systems
(containing $10^5$ electronic spins or more) to relax very quickly
(in $\mu s$). These theoretical insights may have some relevance
to the design of future nanoscopic magnetic devices.

In what follows we shall analyse the role of electrons (Section \ref{sec:2}),
phonons (Section \ref{sec:3}), and nuclei (Section \ref{sec:4}), acting
in isolation on $\vec{S}$. Then in Section \ref{sec:5} we shall put
them together, to give the final rather complex picture. As noted above, we
save detailed discussion of
experiments for other papers; but in  Section \ref{sec:6} we will indicate
the general nature of our predictions.

\subsection{The Model, and Energy Scales}

We briefly describe the giant spin model here, and the various energy scales
relevant to the physics. The basic idea behind this model has been
explained in detail elsewhere \cite{9,11}. The exchange couplings $J_{ij}$
between electronic spins in the grain are enormous compared to the
anisotropy couplings $K_{\alpha}$ (typically $J_{ij} \sim \:eV$, whereas
$K_{\alpha} \sim 10^{-2}-1\: K$). The mesoscopic grains can be treated  as
giant rotators, keeping only states $\mid S,m \rangle $, with
$S \ge m \ge -S$. The spectrum of this giant spin is shown in Fig.1 in a small
bias field $\xi_H$. Sometimes, as a model example, we shall use the
easy-axis/easy-plane system, having Hamiltonian
\begin{equation}
H_o^{\perp}( {\vec S} ) ={1 \over S} \bigg[ -K_{\parallel}\:S_z^2 +
K_{\perp}S_y^2 \bigg] -\gamma \vec{S} \cdot \vec{H}_o \;.
\label{1.1}
\end{equation}
We will usually assume $\vec{H}_o =\hat{\vec{z}} H_o$, so $\xi_H =\gamma SH_o$.

The low energy physics of (\ref{1.1}) can be understood in terms of the
truncated Hamiltonian
\begin{equation}
 H_o( {\vec \tau } ) = (2 \Delta_o  {\hat \tau }_x \cos \pi S -\xi_H
{\hat \tau }_z )\equiv \Delta_s {\hat \tau }_x -\xi_H
{\hat \tau }_z\;,
\label{1.2}
\end{equation}
where the tunneling splitting $\Delta_o$ is
\begin{equation}
 \Delta_o \sim \Omega_o e^{-A_o} \;,
\label{1.3}
\end{equation}
and the bare parameters are
\begin{equation}
\Omega_o \sim  2  (K_{\parallel}K_{\perp})^{1/2}  \;.
\label{1.4}
\end{equation}
\begin{equation}
 A_o \sim 2S(K_{\parallel}/K_{\perp})^{1/2} \;,
\label{1.5}
\end{equation}
 The "bounce frequency" $\Omega_o$ is also roughly the distance to the
next pair of excited levels; the total barrier height in (\ref{1.1}), between
the 2 semiclassical minima at ${\vec S} = \pm \hat{\vec{z}} S$, is
$S K_{\parallel}$. Thus if, say, $S=10^4$, we might easily have
$\Omega_o \sim 0.1-1\:K$, a barrier height $\sim 0.01-0.1\: eV$, but
a splitting $\Delta_o \sim 10^{-6}\:K$, or less.

This is certainly not the whole story. Magnetic grains having more complicated
symmetries are not so easily described using WKB/instanton methods.
The "internal" magnon modes of the giant spin
are also neglected, except for those uniform precession modes corresponding
to $\mid S,m \rangle $.
There will also be phonons (spectrally very weak at these low energies) and
possibly electrons (which will mix strongly with all states of the grain).

Finally and most importantly, the nuclei in the grain (and also possibly
outside) will couple via hyperfine interactions to each level shown in Fig.2.
The net result of this is shown in Fig.3 for the 2 lowest levels in the
presence of a bias field. Typically $\Omega_o \gg \omega_k \gg \Delta_o$,
where $\omega_k $ is an {\it individual} hyperfine coupling;
however $N$ nuclei will spread out a single grain level into a mass of
$2^N$ levels, of Gaussian half-width $\sim N^{1/2} \omega_o$, where
$\omega_o$ is the principal hyperfine splitting. For rare earths, where
$\omega_o \sim 1\:GHz$ or greater, this half-width can easily exceed
$\Omega_o$ for mesoscopic grains, and the whole
"giant spin" model begins to get rather complex.

In this paper we will ignore all physics at energies $\sim \Omega_o$
or greater, simply including it into renormalisation parameters of a low-$T$
effective Hamiltonian. For such a treatment to be valid, we then require
\begin{equation}
 k_BT,\;\xi_H,\;N^{1/2} \omega_o \ll \Omega_o  \;.
\label{1.6}
\end{equation}

We will also assume that any "loose spins" on the surface of the grain
still have couplings $\gg \Omega_o$ to $\vec{S}$.
If not, they will be treated as part of the "spin bath" environment, along
with nuclei and paramagnetic
impurity spins.

Before beginning, it is useful to bear in mind 2 simple points.
Note first that in the presence of any bias
$\xi_H$ which is greater than $\Delta_s$ in (\ref{1.2}), the
system is essentially "frozen" in one or other of the states
$\mid \Uparrow \rangle $ or $\mid \Downarrow \rangle $. Suppose, e.g.,
the system starts in the higher-energy state $\mid \Uparrow \rangle $ at
$t=0$. Then the probability $P^{(0)}(t,\xi_H)$ of finding it
$\mid \Uparrow \rangle $ at time $t$, in the absence of any other
couplings, is given by
\begin{equation}
 P^{(0)}(t,\xi_H) = 1 -{ \Delta_s^2 \over E^2} \sin ^2 Et  \;,
\label{1.7}
\end{equation}
where $E^2 = \xi_H^2+\Delta_s^2$. Thus the coupling to phonons, electrons, and
nuclei is necessary if we are to have any relaxation at all. Usually one thinks
 of this as a kind of spontaneous or stimulated emission, perhaps in the
presence of dissipation, in which the requisite energy is taken up by a
bath of oscillators.

In fact this conventional picture of tunneling of a biased 2-level system,
coupled dissipatively to a bath of oscillators \cite{12}, works
reasonably well for a giant spin coupled to either electrons or phonons.
This we shall see in detail in the next two sections. However as soon as we
introduce the spin bath of nuclei, etc., it breakes down completely; this
is basically why our final results are so surprising.

The second point arises from the finite-dimensional nature of the giant spin
Hilbert space. As emphasized by van Hemmen and Suto \cite{4,13}
this means that we must use WKB  calculations of tunneling rates
with great care when discussing the motion of $\vec{S}$. This point
is especially clear in recent work of Politi et al. \cite{14} (see
also Villain et al. \cite{14}), analysing the Mn$_{12}$O$_{12}$
acetate systems; here tunneling appears to proceed via a 4th-order
term $\sim (\hat{S}_+^4 +\hat{S}_-^4) $ in $H_o(\vec{S} )$. Consequently
transitions may only proceed between states differing by $\Delta S_z =\pm 4$.
To correctly handle such selection rules in the instanton calculus
is not a trivial problem. Even more important, there is no
continuum of final states in this tunneling problem, and tunneling may
only proceed if there is near resonance between initial and final states.
This means that in the presence of a bath of, e.g., phonons, one may {\it not}
in general simply apply the Caldeira-Leggett tunneling formalism \cite{15}
to calculate tunneling, since this formalism assumes a continuum of final
states. For this reason calculations such as those of Garg and Kim \cite{16},
of the tunneling rate of a strongly biased grain interacting with phonons,
can only be correct when $S \to \infty$. For finite $S$ one must take account
of the discrete nature of the spectrum of $S$, at least when the level spacing
is greater than the (very small) linewidth caused by phonon damping (note
further that calculations of grain tunneling in the presence of a nuclear spin
bath should {\it never} use the Caldeira-Leggett method, since the spin bath
cannot be mapped onto an oscillator bath of the Feynman-Vernon \cite{17}
type - thus calculations of this kind \cite{18} are not valid for any $S$).

We now turn to the detailed treatment of the different relaxation mechanisms.
Readers wishing to see only a summary of the results should proceed directly
to section \ref{sec:6}.

\section{Giant Spins and the Electron Bath}
\label{sec:2}

In this section we develop a model for the coupling between the giant spin
and the electron bath only, and use it to calculate relaxation rates for the
grain magnetisation. Since, as far as
we are aware, no attempt has ever been made to deal with
the coupled giant spin/electron system  before,
it is clear that we must begin from first principles. The full development
of our model, including detailed discussion of the multiple-scattering
of the electrons by the individual spins inside the grain, the effects due to
surface spins, to the discreteness of the electronic spectrum inside the grain
if it is on an insulating substrate, the interaction between electrons and the
internal degrees of freedom (magnons) in the grain, etc., will not appear in
this paper. We shall also set aside questions about how conduction electron
dynamics {\it outside} the grain, in a conducting substrate, will be influenced
by the giant spin dynamics. This is a very important question for any
experiment (which can use the substrate magnetoresistance as a probe of the
grain dynamics), but is too complicated to deal with here.

There are 3 physical situations which we will address here.
The first involves conducting electrons both inside and outside the
grain; we shall see that this problem can be analysed fairly clearly. Likewise
the problem of an insulating grain interacting with conducting substrate
can be modelled in a reasonably treatable way. The problem of a conducting
grain
and an insulating substrate is more messy, however we can fairly easily
extract the essential results. Our essential result is that unless the
substrate is insulating, electrons simply freeze $\vec{S}$
, unless $S$ is very small. We emphasize that
in reality, nuclear spins must be included (Sections \ref{sec:4} and
\ref{sec:5}).

\subsection{Conducting Grain and Conducting Substrate.}

We shall start by assuming that the giant spin, contained within a volume
$V_o \sim R_o^3$, interacts with an electron fluid  which permeates freely
through the boundary between grain and substrate; i.e., a Hamiltonian
\begin{equation}
H=H_o( {\vec S} ) +\sum_{\vec{k},\sigma} c^{\dag}_{\vec{k},\sigma}
c_{\vec{k},\sigma} +{1 \over 2}  \sum_{{\bf i}\in V_o} J_{\bf i}
\hat{\vec{s}}_{\bf i}\cdot \hat{\vec{\sigma}}^{\alpha \beta}
 \sum_{\vec{k}}\sum_{\vec{q}} e^{i\vec{q}\cdot \vec{r}_{\bf i}}
c^{\dag}_{\vec{k}+\vec{q},\alpha}
c_{\vec{k},\beta} \;,
\label{2.1}
\end{equation}
describing short-range exchange interaction between the localised individual
spins $\vec{s}_{\bf i}$, located at positions $r_{\bf i}$ inside the grain,
and conduction electrons in momentum $\mid \vec{k} \rangle $. A more
refined model would couple grain electrons
in eigenstates $\mid \mu \rangle $ to substrate electrons
$\mid \vec{k} \rangle $, via a transfer matrix $\hat{T}_{\mu \vec{k}}$ across
the boundary. However it is intuitively obvious that this is unnecessary
once $\mid \hat{T}_{\mu \vec{k}} \mid > \Delta \epsilon_{\mu }$, where
$\Delta \epsilon_{\mu }$ is the typical level spacing between internal electron
states of the grain. If the electronic bandwidth is $D$, then
$\Delta \epsilon_{\mu } \sim D/S$, and for most metals $D \sim 1-5\:eV$; thus
if $S \sim 10^3$, $\Delta \epsilon_{\mu } \sim 10-100\:K$, so that an
insulating layer at least $10 \:\AA$ thick would be required to make such
a model necessary.

We assume in using (\ref{2.1}) that the internal grain spins are locked
together to form a giant spin, and that the $J_{\bf i}$ already take
account of the virtual mixing between internal excited states of the
magnetic ions and of the giant spin (magnons, surface modes) and the
electron bath.
Thus, as usual, we shall be working at low $T$ and small  $\vec{H}_o$.
However we shall assume that the $J_{\bf i}$ can vary; near the
surface of the grain boundary they may be a little weaker.
The present  model
does not include "loose spins", i.e., surface spins which are
more weakly coupled to the motion of $\vec{S}$ (the reader should not
confuse the Kondo coupling $J_{\bf i}$ here with the inter-spin exchange
$J_{ij}$ between spins inside the grain). These will be discussed
in section \ref{sec:4} onwards. Typically the Kondo couplings
$J_{\bf i} \sim 0.1-1\: eV$.

We now rewrite (\ref{2.1}), incorporating the giant spin hypothesis,
to get a volume averaged interaction
\begin{equation}
H_{int}^b \to  {1 \over 2} \overline{J}
\vec{S}\cdot \hat{\vec{\sigma}}^{\alpha \beta}
 \sum_{\vec{k}}\sum_{\vec{q}} F_{\vec{q}}
c^{\dag}_{\vec{k}+\vec{q},\alpha}
c_{\vec{k},\beta} \;;
\label{2.2}
\end{equation}
\begin{equation}
 F_{\vec{q}}=\int_G { d^3\vec{r} \over V_o }
 \:\rho (\vec{r}) e^{i\vec{q}\cdot \vec{r}} \;,
\label{2.3}
\end{equation}
where the form factor $F_{\vec{q}}$ integrates the number density
$\rho (\vec{r})$ of microscopic spins over the grain volume, and
$\overline{J}$ is the mean value of the $J_{\bf i}$. $H_{int}^b$ is a sort
of "giant spin Kondo Hamiltonian", and as such contains a lot of interesting
physics, much of which we will ignore in this paper. Recall that the Kondo
coupling for microscopic spins is just (\ref{2.2}) with
$\rho (\vec{r})$ in (\ref{2.3}) restricted to a single lattice cell; for
this problem one conventionally defines a dimensionless coupling
$g=\overline{J} N(0) \sim 0.1$ for most metals, where $N(0) \sim \rho /D$
is the Fermi surface density of states. The Kondo problem  can be mapped
onto an Ohmic spin-boson problem of the kind discussed by Leggett {\it et al}.
\cite{12}; even for this spin-$1/2$ system, there is a large dimensionless
Ohmic coupling to the electron bath; $\alpha_K \sim O(1)$.

It is possible to start with (\ref{2.1}) and build up an analogous description
by considering the individual Kondo scattering from each spin, in many
different
orbital channels. Here we shall finesse entirely this move by instead
integrating out the electrons directly, starting from (\ref{2.2}). We
first truncate directly to a biased spin-boson Hamiltonian of form
(for $H_o(\vec{S}) = H_o^{\perp}(\vec{S})$ in (\ref{1.1})):
\begin{equation}
H_{eff}^b =\Delta_o \hat{\tau}_x  - \epsilon \hat{\tau}_z +
 {1 \over 2} \sum_{k=1}^N m_k(\dot{x}_k^2 + \omega_k^2 x_k^2) +
\hat{\tau}_z \sum_{k=1}^N c_k x_k \;,
\label{2.4}
\end{equation}
where the bias field is parallel to the easy axis (we shall ignore transverse
fields here). The Caldeira-Leggett spectral function for this problem
is
\begin{equation}
J_b(\omega )= {\pi \over 2} \sum_{k=1}^N {\mid c_k \mid^2 \over
m_k \omega_k } \equiv \pi \alpha _b \omega \;,
\label{2.5}
\end{equation}
and the Ohmic coupling constant $\alpha_b$ is determined by a standard
Fermi surface average \cite{19,20} over electrons, i.e.,
\begin{equation}
\alpha_b = 2g^2S^2 \langle \mid
F_{\vec{k}-\vec{k}^\prime } \mid ^2 \rangle _{F.S.} \;.
\label{2.6}
\end{equation}
The naive guess that $\alpha_b \sim S^2$, apparently confirmed by
(\ref{2.6}), is however wrong; carrying out the average, one gets
\begin{eqnarray}
\alpha_b & = & 2g^2S^2 \int_G {d^3\vec{r}d^3\vec{r}^\prime \over V_o^2}
 \left(  {\sin k_F \mid \vec{r} - \vec{r}^\prime \mid \over k_F
\mid \vec{r} - \vec{r}^\prime \mid } \right)^2 \nonumber \\
& \sim & \left( {gS \over k_FR_o }  \right)^2 \sim g^2S^{4/3} \;,
\label{2.7}
\end{eqnarray}

This at first surprising result is a consequence of interference between
Kondo scattering at different sites - notice that $k_F^{-1} \ll R_o$, so that
the problem is a multiple scattering one. A related point is that the estimate
of $\alpha_b$ in (\ref{2.7}) looks  perturbative -
it is not clear it should work for large $\alpha_b$.
However for a grain of radius $R_o$ some
$l_{max}^2 \sim 1/(k_FR_o)^2 \sim S^{2/3}$ orbital channels are playing role
in electron scattering. If $S \le 10^3$,
the phase shifts in each channel are small
(one has $\delta_l \sim g k_FR_o \sim gS^{1/3}$),
making it possible to apply a
perturbation theory, yet the sum over all the channels
can be rather large. The case of large phase shifts is more complicated,
and may result in a giant spin dynamics  different from that
described by the standard Ohmic model.
We address this possibility in Appendix A.

For mesoscopic spins ($S \ge 10^3$) the coupling in (\ref{2.7}) is very large;
we may take over standard results from the spin-boson problem to discuss
its effect on the dynamics of $\vec{S}$. Since $\alpha_b >1$, coherence
at zero bias is destroyed at any temperature; at $T=0$ the grain dynamics
are rigidly locked, and the spin $\vec{S}$ cannot move. At finite $T$ and
$\epsilon =0$, the system fluctuates incoherently between the states
$\mid \Uparrow \rangle$ and $\mid \Downarrow \rangle$ at a rate $\sim
(T/\Omega_o)^{2\alpha_b -1}$; on the other hand if $\epsilon \gg T$, one
has incoherent relaxation at a rate $\sim (\epsilon /\Omega_o)^{2\alpha_b -1}$.
The general result for the relaxation rate is \cite{12,20}
\begin{equation}
\tau_e^{-1} = 2\Delta_o \left( {\Delta_o \over \Omega_o } \right)
\left( {2 \pi T \over \Omega_o } \right)^{2\alpha_b-1}  \cosh
\left( {\epsilon \over 2  T } \right)
{ \mid \Gamma [\alpha_b +i\epsilon /2\pi T ] \mid^2 \over \Gamma [2\alpha_b] }
\;,
\label{2.8}
\end{equation}
 where $\Gamma [z]$ is the Gamma function; for $\epsilon =0 $, this gives
\begin{equation}
\tau_e^{-1}(T) = \Delta_o \left( {\Delta_o \over \Omega_o } \right)
{ \Gamma^2 [\alpha_b  ] \over \Gamma [2\alpha_b] }
\left( {2 \pi T \over \Omega_o } \right)^{2\alpha_b-1} \;.
\label{2.9}
\end{equation}
On the other hand for very low $T$, such that $\epsilon /k_BT \gg 1$, it gives
\begin{equation}
\tau_e^{-1}(\epsilon ) =
2 \pi \Delta_o \left( {\Delta_o \over \Omega_o } \right)
{ 1 \over \Gamma [2\alpha_b] }
\left( {\epsilon  \over \Omega_o } \right)^{2\alpha_b-1} \;.
\label{2.10}
\end{equation}

The crucial thing to notice here is that for microscopic spins ($S \ll 100$),
there will be easily observable relaxation of $\vec{S}$. On the
other hand if $S$ is much greater than $100$, the giant spin will be
frozen completely, even over astronomical times, unless either bias
or temperature is large ($\epsilon , k_BT \sim \Omega_o$). Moreover, in this
model we would expect that applying either a large bias or temperature
would lead to a sudden "switching on" of the relaxation at some critical bias
or temperature (however recall that our model is not strictly valid for
$\epsilon , k_BT \sim \Omega_o$, since higher levels of $\vec{S}$
will then enter into the picture).

To see this, consider 2 examples. First, suppose that
$g^2 \sim 10^{-2}$ and $S \sim 30$, so $\alpha_b \approx 1$. Then (\ref{2.9})
shows that for bias $\epsilon =0$, we have $\tau^{-1}(T) \sim \Delta_o^2 T
/\Omega_o^2$, and if $k_BT =0$, we have $\tau^{-1}(T) \sim \Delta_o^2 \epsilon
/\Omega_o^2$. Since $ \Delta_o^2
/\Omega_o$ might be something like $100\: Hz$, this implies low $T$ or low
$\epsilon $ relaxation times of fractions of a second.

On the other hand, consider a second example where
$g^2 \sim 10^{-2}$ but $S \sim 300$, so $\alpha_b \approx 20$.
Then if $\epsilon =0$, we find that $\tau^{-1} (T) \sim 10^{-12}\;
 \Delta_o^2 /\Omega_o \; (2\pi k_BT /\Omega_o)^{40}$, and if $T=0$, we
find $\tau^{-1} (\epsilon ) \sim 10^{-46}\;
 \Delta_o^2 /\Omega_o \; (\epsilon  /\Omega_o)^{40}$.Then if $\epsilon =0$
we see that the relaxation suddenly switches on at a temperature
$k_BT \sim \Omega_o/2 \pi $ (if we are interested in a relaxation time of
order seconds); for temperatures  below this, relaxation times
increase catastrophically (if $T \sim T_c/2$, we find $\tau^{-1} (T) \sim
10^6 $ years!). The same happens as a function of bias when $T \ll \epsilon$.
Obviously if we go to even larger grains, with $S\ge 1000$, this switching
process becomes almost a step function of temperature. Although our
2-level model starts to break down once $\pi k_BT, \epsilon \sim \Omega_o$,
this will not affect the validity of this switching result, or our
calculation of the switching temperature.

We have not included nuclear spins in these calculations - in section
\ref{sec:5} we shall see how a combination of nuclear spins and electrons
affects the dynamics of $\vec{S}$.

\subsection{Insulating Grain and Conducting Substrate.}

The insulating grain can be analysed in a similar way, provided we gloss over
some of the complications arising from the variation of both the electron
density and the spin properties in the vicinity of the surface. Thus
we naively extend the previous model by writing an effective interaction
Hamiltonian
\begin{equation}
H_{int}^s \sim  {1 \over 2} \overline{J}
\vec{S}\cdot \hat{\vec{\sigma}}^{\alpha \beta}
 \sum_{\vec{k}}\sum_{\vec{q}} F_{\vec{q}}^s
c^{\dag}_{\vec{k}+\vec{q},\alpha}
c_{\vec{k},\beta} \;;
\label{2.11}
\end{equation}
\begin{equation}
 F_{\vec{q}}^s=\int_G { d^3\vec{r} \over V_o }
 \:\mid \Psi_e(\vec{r}) \mid^2\rho (\vec{r}) e^{i\vec{q}\cdot \vec{r}} \;,
\label{2.12}
\end{equation}
where now the form factor takes account of the decay of the electron density
$\mid \Psi_e(\vec{r}) \mid^2$ as one penetrates into the sample. By the same
manoeuvres as before this leads to an Ohmic coupling
\begin{equation}
\alpha_s \sim g_s^2 S^{2/3} \;,
\label{2.13}
\end{equation}
where $g_s \sim g$, but is multiplied by some "scale factor" which describes
the depth to which the electrons penetrate into the grain;
 clearly it can vary widely.

The reduction to only near surface spins makes $\alpha_s \ll \alpha_b$ for
large $S$, but from (\ref{2.13}) we see that $\alpha_s \ge O(1)$ if
$S  \ge 10^{3}$. For smaller grains, i.e., $S \le 100$, one can have
$\alpha_s <1$, and then results different from the "frozen spin" behaviour
of (\ref{2.8})-(\ref{2.10}) are obtained. We may define a renormalised
splitting $\Delta_s$, and a damping rate $\Gamma_s$, by
\begin{equation}
\Delta_s =\Delta_o (\Delta_o /\Omega_o )^{\alpha_s/(1-\alpha_s)}\;;
\label{2.14}
\end{equation}
\begin{equation}
\Gamma_s = 2 \pi \alpha_s T\;.
\label{2.15}
\end{equation}

At $T=0$ one can get \cite{4} coherent motion if $\alpha_s < 1/2$; however
 any such coherence is destroyed once $\Gamma_s \ge \Delta_s$, which
even for microscopic values of $S$ (i.e., $S \sim 10$) will occur at
extremely low temperatures, usually much less then $1\:mK$. For higher
$T$ we find, when $\epsilon =0$ and $\alpha_s <1$, that $\tau^{-1}$ is still
given by (\ref{2.9}), which in terms of $\Delta_s$ is
\begin{equation}
\tau_e^{-1}(T) = 2 \Delta_s
{ \Gamma^2 [\alpha_s  ] \over \Gamma [2\alpha_s] }
\left( {2 \pi T \over \Delta_s } \right)^{2\alpha_s-1} \;.
\label{2.16}
\end{equation}
For finite $\epsilon$ and $T$, the form (\ref{2.8}) is still valid, as is
(\ref{2.10}) when $\epsilon \gg T$.

{}From all of these results we  conclude that for low $T$ and $\epsilon$,
the relaxation dynamics of mesoscopic grains will be extremely slow in
the presence of a conducting substrate.
Thus, although our models have
certainly not included all possible nuances of the surface physics, one
conclusion is very clear and unlikely to be affected by further refinements,
i.e., that for $T, \epsilon \ll \Omega_o$, mesoscopic magnetic grains on
conducting substrates will have their spin dynamics completely frozen.
However once either $T$ or $\epsilon$ is of order $\Omega_o$ or greater,
these dynamics will rapidly be liberated. Again, these statements will be
modified if nuclear spin effects are important (Section \ref{sec:5}).

\subsection{Conducting Grain and Insulating Substrate.}

In the opposite extreme when the substrate is insulating, the character
of the electron states is quite different - they form a set of discrete
states $\mid \mu , \vec{\sigma} \rangle $ inside the grain, with mean
spacing $\Delta \epsilon_\mu \sim D/S $ for bandwidth $D$. We describe these
states via creation and annihilation operators
$A^{\dag}_{\mu ,\sigma}$,$A_{\mu , \sigma}$; these states will be
rather complicated, even in the absence of spin-orbit coupling. It is very
useful to separate the exchange scattering interaction into  bulk and
surface parts, where now the bulk part sums over the whole sample, assuming
that all the $J_{\bf i}$ are the same, whilst the surface part describes
the {\it difference } between $H_{int}^{bulk}$ and the true interaction
Hamiltonian $H_{int}$; thus if
\begin{equation}
H_{int}={1 \over 2}   \sum_{{\bf i}\in V_o} J_{\bf i}
(\hat{\vec{s}}_{\bf i}\cdot \hat{\vec{\sigma}}^{\alpha \beta})
c^{\dag}_{\bf i,\alpha}
c_{\bf i,\beta} \;,
\label{2.17}
\end{equation}
with the $c^{\dag}_{\bf i,\alpha}$ creating electron on site ${\bf i}$, we
write
\begin{eqnarray}
H_{int}^{bulk} &= & {1 \over 2} \overline{J}
\sum_{{\bf i}\in V_o}
(\hat{\vec{s}}_{\bf i}\cdot \hat{\vec{\sigma}}^{\alpha \beta} )
c^{\dag}_{\bf i,\alpha} c_{\bf i,\beta} \nonumber \\
& = & { \overline{J} \over 2S}
(\vec{S}\cdot \hat{\vec{\sigma}}^{\alpha \beta})
\sum_{{\bf i}\in V_o} c^{\dag}_{\bf i,\alpha} c_{\bf i,\beta} \equiv
{\overline{J} \over 2S}
(\vec{S}\cdot \hat{\vec{\sigma}}^{\alpha \beta})
\sum_{\mu} A^{\dag}_{\mu ,\alpha }A_{\mu , \beta} \;,
\label{2.18}
\end{eqnarray}
and the surface term
\begin{equation}
H_{int}^S={1 \over 2}  \sum_{{\bf i}\in V_o} (J_{\bf i}- \overline{J})
(\hat{\vec{s}}_{\bf i}\cdot \hat{\vec{\sigma}}^{\alpha \beta})
c^{\dag}_{\bf i,\alpha} c_{\bf i,\beta} \;.
\label{2.19}
\end{equation}

We notice that the bulk term is diagonal in both the site representation
and the exact eigenstate representation, because of completeness of states.
This
is important, because the condition $\overline{J} \gg \Omega_o$ means
that the electron spins inside the system will rotate rigidly with $\vec{S}$,
and from the last form in (\ref{2.18}) we see that there is no way this
adiabatic rotation can excite the low energy orbital states $\mid \mu , \sigma
\rangle $ of the system. Thus the effect of $H_{int}^{bulk}$ is simply
to renormalise the moment of inertia of the giant spin.

On the other hand the surface term does have off-diagonal matrix elements
$\langle \mu \mid H_{int}^S \mid \mu^\prime \rangle $ and  this may result
in  infrared electron-hole pair production when $\vec{S}$ flips.
The calculation of the corresponding  coupling  parameter
$\alpha_s$ proceeds in exactly the
same way as in the previous subsection (the corrections to the plane-wave
calculation are small in a large grain with $k_FR_o \gg 1$). Thus
at high temperatures the  internal conduction electrons constitute an Ohmic
bath with
\begin{equation}
\alpha_s \sim g_s^2 S^{2/3} \;,
\label{2.20}
\end{equation}
(here $g_s$ is given by $(J_{\bf i}- \overline{J})N(0)$
averaged over the surface).
Now we notice that this result holds only at $T> \Delta \epsilon_{\mu }$.
At lower temperature the effective coupling constant goes to zero exponentially
with $T$:
\begin{equation}
\alpha_s^{eff} \sim  \alpha_s e^{-\Delta \epsilon_{\mu }/T} ;,
\label{2.21}
\end{equation}
and the infrared renormalisation of the tunneling rate stops at
\begin{equation}
\Delta_s = \Delta_o \left( {\Delta \epsilon_{\mu } \over \Omega_o }
\right)^{\alpha_s} ;, \;\;\;\;(for \;\; \Delta \epsilon_{\mu } \ll \Omega_o
)\;.
\label{2.22}
\end{equation}
If $\Delta \epsilon_{\mu } \gg \Omega_o$  one has $ \Delta_s \approx \Delta_o$,
and we do not have to worry about conduction electrons in the grain at all
because they will rotate adiabatically with $\vec{S}$.

The onset of coherent motion in zero bias can be found by comparing
the damping rate $\Gamma_s = 2 \pi \alpha_s^{eff} T$ with $\Delta_s$.
Expressing all the parameters in terms of $S$ we find the criterion of
coherence in the form
\begin{equation}
{T \over \Delta \epsilon_{\mu } } \sim S {T \over D } \ll
{1 \over \ln {2\pi g^2S^{2/3} D \over S\Delta_s}}\;.
\label{2.23}
\end{equation}
For experiments at $mK$ temperatures and small bias
$\epsilon < \Delta \epsilon_{\mu }$, this means that
electronic dissipation effects can be ignored in such grains
provided $S \le 10^5$. On the other hand for $T$ and/or
$\epsilon > \Delta \epsilon_{\mu }$, we return to the case of
strong Ohmic dissipation which looks very much the same as that for
an insulating grain on a metallic substrate.

\section{Giant Spins and the Phonon Bath }
\label{sec:3}

In the absence of electrons (and of nuclear spins) relaxation in a biased
system proceeds via phonons. This is a very slow process - typical relaxation
times are of order months. A number of previous studies have appeared;
Garg \& Kim \cite{16} and Chudnovsky \cite{21} have discussed grain tunneling
within a straightforward Caldeira-Leggett approach, assuming a continuum
of final states available for the tunneling - we have already
explained  problems with this approach in the introduction. Politi
{\it et al}. \cite{14} have given a very thorough discussion of the effects
of phonons on tunneling in the Mn$_{12}$O$_{12}$ acetate system, including the
effects of discrete initial and final states, and the non-diagonal couplings
(to be defined below). They also take proper account of the symmetries and
selection rules in the tunneling transitions. Whether their theory explains
the relaxation measurements in Mn$_{12}$O$_{12}$
 acetate \cite{22} is another matter - in our
opinion the explanation may have
more to do with dipolar interactions between the
molecules \cite{23}. The treatment of phonon effects we shall give is
more general although quite close to theirs.
We first discuss the form the low-energy effective
Hamiltonian should take, depending on the symmetries of $H_o(\vec{S})$ and
the direction of the applied bias field, and determine the relaxation
of the giant spin dynamics for the different possible couplings. The values
of the relevant couplings are then estimated - no attempt is made at
exact calculations, since these would depend so much on the detailed
nature of the system in question. Our main result is that these couplings
are so weak that phonons by themselves are incapable of causing
anything but very slow relaxation of $\vec{S}$.

\subsection{Effective Hamiltonians, and Relaxation Rates}

We start by assuming, for ease of discussion, a simple easy-axis, easy-plane
bare giant spin Hamiltonian in a bias (Eqs.(\ref{1.1}), (\ref{1.2})).
The  phonons are
described by $H_{\phi} =\sum_{\vec{q}} \nu_{\vec{q}}
(b^{\dag}_{\vec{q}}b_{\vec{q}}+1/2)$, with $\nu_{\vec{q}} =cq$ (we take
account of polarization later, and ignore optical modes), and the general
form of the truncated interaction is then
\begin{equation}
H_{int} = \hat{\tau}_z \sum_{\vec{q}} C_{\vec{q}}^z(h) x_{\vec{q}} +
\big[ \hat{\tau}_+ {\Delta_o \over \Omega_o }
\sum_{\vec{q}} C_{\vec{q}}^{\perp} x_{\vec{q}}
+H.c.    \big] \;;
\label{3.1}
\end{equation}
\begin{equation}
x_{\vec{q}} = {1 \over (2 m_{\vec{q}} \nu_{\vec{q}})^{1/2}}
(b^{\dag}_{\vec{q}}+b_{-\vec{q}}) \;,
\label{3.2}
\end{equation}
where $h=\gamma_e S H_x$ is the transverse bias.

It is unusual to consider non-diagonal couplings (i.e., couplings in
$\hat{\tau}_\pm $), since in general their effects are reduced, compared
to those of diagonal properties, by a factor
$(\Delta_o /\Omega_o)^2$. However
we shall see that in this system the diagonal couplings are zero unless
we apply a field perpendicular to the easy axis. The reason for this is
very simple - time reversal symmetry implies that the
magnetoacoustic interaction cannot distinguish between grain states
$\vec{S}_1$ and $\vec{S}_2$ when $\vec{S}_1=-\vec{S}_2$ (a point which also
arises in the interaction of phonons with domain walls \cite{6}), and diagonal
couplings describe interactions which do distinguish such states. Applying
a transverse field projects a component of $\vec{S}_1$ on $\vec{S}_2$,
and enables such a coupling; from this argument we see that
$C^z_{\vec{q}} (h)$ is proportional to the transverse field $H_x$. We shall
later see that
\begin{equation}
C^z_{\vec{q}} \sim (\gamma_eH_x /K^{\parallel} ) C_{\vec{q}}^{\perp} \;,
\label{3.3}
\end{equation}
for a field applied in the easy plane.

Before deriving (\ref{3.1}) and evaluating the couplings, we first give the
results for relaxation rates that derive from it. Just as for the electron
bath,
these rates are very low, but for the opposite reason - instead of being frozen
by a very strong coupling, $\vec{S}$ is frozen because the effective
coupling is so weak!

The relaxation rates are derived by first defining Caldeira-Leggett
spectral functions
\begin{equation}
J_{\alpha} (\omega ) = {\pi \over 2} \sum_{\vec{q}} {\mid C_{\vec{q}}^{\alpha}
\mid ^2 \over m_{\vec{q}} \nu_{\vec{q}} } \delta (\omega - \nu_{\vec{q}} ) \;,
\label{3.4}
\end{equation}
where $\alpha =z,\perp$; these describe the phonon effect on grain
dynamics via $2$nd order perturbation theory. It has of course been known for
decades that these couplings are very weak for $\omega \ll \Theta_D$,
the Debye frequency; in fact
\begin{equation}
J_{\alpha} (\omega ) = B_{\alpha} (\omega /\Theta_D )^m\;,
\label{3.5}
\end{equation}
with $m\ge 3$ in $3$ dimensions; this is the fundamental reason for the
coherent motion of polarons and defects in insulators at low temperatures
\cite{20}. We will see below that $m=3$ in our case.

We start with the case where there is no transverse field $H_x$, so that
$C^z_{\vec{q}} =0$ in (\ref{3.1}). The relaxation dynamics for this case
are described as damped  motion between $\vec{S} = \pm \hat{\vec{z}}
S$  with a relaxation rate (see, e.g., \cite{20})
\begin{equation}
\tau^{-1}(\epsilon ) \sim 2 {\Delta_o^2 \over \Omega_o^2 }\;
{ \epsilon ^2 \over E^2}\; J_{\perp} (E ) \coth (\beta E /2) \sim
4 B_{\perp} {\Delta_o^2 \over \Omega_o^2 }\; (\epsilon /\Theta_D )^{2}
{T \over \Theta_D}\;,
\label{3.6}
\end{equation}
\begin{equation}
E^2=\Delta_o^2 + \epsilon^2\;,
\label{3.7}
\end{equation}
where the longitudinal bias $\epsilon = \gamma_e S H_z$. This rate is
of course very small ($\tau^{-1} \ll \Delta_o, \epsilon $).
Notice that $\tau^{-1}(\epsilon )$ is quite different from that rate
which would be
calculated via a naive application of the spin-boson results \cite{4}; in fact
it is smaller by a factor $(\epsilon /\Omega_o )^2$.

On the other hand application of a transverse
field $\vec{H} = \hat{\vec{x}}H_x$ gives
a diagonal coupling, according to (\ref{3.3}), which can be incorporated into
a spin-boson model; in the absence of any bias $\epsilon $ one finds a
relaxation rate
\begin{equation}
\tau^{-1} \approx  2
 J_{z} (\Delta_o ) \coth (\beta \Delta_o /2) \sim
4  B_{z} {\Delta_o \over \Theta_D }^2\; {T \over \Theta_D}\;,
\label{3.8}
\end{equation}
or, in a non-zero bias,
\begin{equation}
\tau^{-1}(h, \epsilon ) \approx  2 {\Delta_o^2 \over E^2}
 J_{z} (E ) \coth (\beta E /2) \sim
4  B_{z}(h)  {\Delta_o \over \Theta_D }^2\; {T \over \Theta_D}\;,
\label{3.9}
\end{equation}
which is independent of $\epsilon$ as long as $\epsilon \ll T$. Using
(\ref{3.3}) we may express $B_z$ in terms of $B_{\perp}$, and write
\begin{equation}
\tau^{-1}(h)  \sim 4 B_{\perp} \left( { \gamma_eH_x \over
 K^{\parallel} } \right) ^2
 {\Delta_o \over \Theta_D }^2  \; {T \over \Theta_D} \;.
\label{3.10}
\end{equation}

These rates are very long, because there are so few phonons available
to absorb the tunneling energy at these frequencies - we shall give
typical numbers once the coupling constant $B_{\perp}$ has been determined.

\subsection{Magnetoacoustic Couplings}

The magnetoacoustic coupling is of course very well understood \cite{1}; its
effect on the WKB tunneling action of a grain has been analysed by
Garg \& Kim \cite{16} and by Chudnovsky \cite{21}. There has been some
confusion about the relation between Chudnovsky's "angular momentum
coupling" and the usual magnetoacoustic coupling, which we try to resolve here.

The essential points we wish to make are that the couplings
$C_{\vec{q}}^{\perp}$, in (\ref{3.1}), and $B_{\perp}$ in (\ref{3.5}), are
given to within dimensionless constants $\sim O(1)$ by
\begin{equation}
\mid C_{\vec{q}}^{\perp} \mid  \sim 2S\Omega_o \mid \vec{q} \mid \;;
\label{3.11}
\end{equation}
\begin{equation}
B_{\perp}  \sim S^2 \Omega_o^2 /\Theta_D \;,
\label{3.12}
\end{equation}
so that the relaxation rate in a longitudinal field $H_z$ is finally
\begin{equation}
\tau^{-1}(\epsilon ) \sim
S^2 \Delta_o {\Delta_o \over \Theta_D }\;
{(\gamma_e SH_z)^3 \over \Theta_D^3 }\;
\coth \left( {\gamma_e SH_z \over 2k_BT } \right) \;,
\label{3.13}
\end{equation}
whereas the relaxation rate in a non-zero transverse field $H_x$ has an
additional term
\begin{equation}
\tau^{-1}_z(h)  \sim
S^2 \Delta_o {\Delta_o \over \Theta_D } \;
\left( { \gamma_e H_x \Omega_o \over \Theta_D K^{\parallel} } \right) ^2\;
{T \over \Theta_D}\;,
\label{3.14}
\end{equation}
By comparing (\ref{3.13}) and (\ref{3.14}) we find their ratio to be
$\sim  ( H_z SK^{\parallel} /
 H_x \Omega_o )^2 $. This ratio is large unless the external field is almost
parallel to $\hat{\vec{x}}$.

To get some ideas of the relaxation rates implied by these formulae, let
us consider a situation where $S=10^3$, $\Delta_o \sim 1\:MHz$, $T \sim
50\:mK$, $\Theta_D \sim 100\: K$, and $\epsilon \sim \gamma_e S H_z \sim
50\: mK$. Then we get $\tau (\epsilon ) \sim 10^4 \: sec$ (i.e., $3$ hours).
Thus, if we ignore nuclear spin effects, the relaxation times at $50\: mK$
are already very long; and $\Delta_o  \sim 1\:MHz$ is actually a large value
for $\Delta_o$. If the bias energy is further reduced to the $\Delta_o$ scale
we get astronomically large $\tau$.

To get (\ref{3.11}) and  (\ref{3.12}) we start by noting that Chudnovsky's
angular momentum coupling
\begin{equation}
H_{AM} = \int_G d\vec{r} { \dot{\vec{S}} \over S} \cdot (\nabla \times
\vec{u} (\vec{r}) ) \;,
\label{3.15}
\end{equation}
integrated over the grain volume, is nothing but that set of transverse terms
in the magnetoacoustic coupling which is responsible for transporting
angular momentum to the transverse phonons. This magnetoacoustic coupling
has the general form
\begin{equation}
H_{ME} = (\gamma_e /S)^2 \int_G d\vec{r}
a_{ijkl} \vec{S}_i \vec{S}_j u_{kl} (\vec{r})\;,
\label{3.16}
\end{equation}
where the coefficients $\gamma_e^2 a_{ijkl}$ are energy densities, and in fact
are simply combinations of the magnetic anisotropy constants (i.e., of
$K^{\parallel}$ and $K_{\perp}$ in the case of $H_o^{\perp}(\vec{S})$), since
this is the only energy scale that can be involved. From this we see that
for the grain, $H_{ME} \sim SK \mid  \vec{q} \mid $.
However we also have
\begin{equation}
\dot{\vec{S}} = i  [\vec{S},\: H_o(\vec{S}) ] \sim \Omega_o S \;,
\label{3.17}
\end{equation}
and so $H_{AM} \sim S \Omega_o \mid  \vec{q} \mid $ also; the direct
relation to the $a_{ijkl}$ can be easily determined for any particular
$ H_o(\vec{S})$ and lattice type.

{}From this it follows that $ C_{\vec{q}}^{\perp}
 \sim 2S\Omega_o \mid \vec{q} \mid $ in (\ref{3.1}).
It is instructive to rederive this by estimating the correction $\delta A$ to
the bare tunneling action in a given strain field $u_{kl}$, i.e., finding
\begin{equation}
\Delta_o(\{ u_{kl} \} ) = \Omega_o e^{-A(\{ u_{kl} \} ) }
\sim \Delta_o e^{-\delta A(\{ u_{kl} \} ) } \;,
\label{3.18}
\end{equation}
This is determined trivially from the interaction $H_{AM}$ by integrating
over time (assuming a negligible change  in the instanton path); writing
\begin{equation}
 \delta A(\{ u_{kl} \} ) \equiv
\sum_{\vec{q}} { C_{\vec{q}}^{\perp} \over \Omega_o } x_{\vec{q}} \;,
\label{3.19}
\end{equation}
we get
\begin{eqnarray}
 C_{\vec{q}}^{\perp} &=& 2S\Omega_o \int_G {d\vec{r} \over V_o }
e^{i\vec{q} \cdot \vec{r}} (\vec{q} \times \hat{\vec{e}}_{\vec{q}})\cdot
\hat{\vec{z}} \nonumber \\
&=& 2S\Omega_o F_{\vec{q}} (\vec{q} \times \hat{\vec{e}}_{\vec{q}})\cdot
\hat{\vec{z}} \;,
\label{3.20}
\end{eqnarray}
where $ F_{\vec{q}}$ is the form factor, and $\hat{\vec{e}}_{\vec{q}}$ is
the polarization of the relevant phonon; for long wavelength
 phonons ($qR_o \ll 1$) this gives us equation (\ref{3.11}); and
$B_{\perp}$ in (\ref{3.12}) immediately follows using (\ref{3.4}). Notice that
in the case of the easy-axis/easy-plane Hamiltonian $H_o^{\perp} (\vec{S})$,
the same argument gives
\begin{equation}
 C_{\vec{q}}^{z} \sim 2S \Omega_o { \gamma_e H_x \over K^{\parallel} }
\mid \vec{q} \mid  \;.
\label{3.21}
\end{equation}

It is sometimes useful to write all these couplings in terms of the relevant
sound velocity, using $\Theta_D^4 \sim \rho c^5$; usually this will
mean the transverse sound velocity since it is usually lower than the
longitudinal one. Then without specifying exactly which of the many
combinations of the $a_{ijkl}$ enter into our expression for some particular
symmetry, but just calling this combination $K_a$, we write
 \begin{equation}
J_{\perp} (\omega ) \sim S^2 { K_a^2 \over \rho c^5} \omega^3 \;;
\label{3.22}
\end{equation}
 \begin{equation}
J_{z} (\omega ) \sim { K_a^2 \over \rho c^5} \left(
{ MH_x \over K^{\parallel} } \right)^2 \omega^3 \;.
\label{3.23}
\end{equation}

Finally, we remind the reader that none of the analysis in this section takes
 account of nuclear spins.

\section{Giant Spins and the Spin Bath}
\label{sec:4}

We now turn to a very different kind of bath from the phonon and electron bath,
viz., the "spin bath" made up from nuclear spins both inside and outside the
grain, paramagnetic electronic impurity spins, and any loose spins in or
near the grain surface. As noted in the introduction, this spin bath cannot be
mapped onto an oscillator bath, and so a completely different kind
of theoretical framework is necessary to describe its effects on mesoscopic
or macroscopic quantum objects.

The main point of this section is to show how the spin bath, interacting in
isolation with a giant spin $\vec{S}$, will basically block
any relaxation at all, unless either the grain happens to be in a total
field (produced by the sum of the external field and the "internal
bias field" generated by the nuclei) which is very small, less than roughly
$\Delta_o$; {\it or } the nuclear spin diffusion mechanism of energy bias
diffusion allows the system to find the resonance window.
 In what follows we first briefly recall the 4 mechanisms by which the spin
bath controls the giant
spin dynamics, and set up the formalism required to analyse magnetic relaxation
in the presence of nuclear spins. We then derive the
relaxation dynamics of $\vec{S}$, first ignoring the internuclear spin
diffusion (to see how $\vec{S}$ is frozen), then including it (to
show how $\vec{S}$ can then relax).

\subsection{Effective Hamiltonian, and the 4 Coherence-Blocking Mechanisms.}

In an earlier  series of papers \cite{10,23,24,25,26}
we have constructed a theoretical description of
a giant spin interacting with a more or less arbitrary spin bath, and solved
this model for the quantum dynamics of $\vec{S}$ in the absence of any external
bias. The same sort of model has also been set up to describe an
arbitrary unbiased macroscopic coordinate interacting with a spin bath.
Elsewhere we have used this model to deal with coherence experiments in
SQUID's \cite{27}, and also shown how it leads to some remarkable conclusions
about the quantum theory of measurement (particularly concerning the role that
nuclear spins play in making the mesoscopic world behave classically
\cite{28}).

Here we generalise this work to deal with a {\it biased} giant spin. As before,
we shall assume that the giant spin can be truncated to 2 levels, and that
$T, \epsilon \ll \Omega_o$; all effects of higher levels will be absorbed
into the parameters of an effective Hamiltonian. This Hamiltonian was derived
in Refs.\cite{10,24,25,26}, and is written for the biased case as
\begin{equation}
 H_{eff}=2 \Delta_o \big\{ {\hat \tau }_+\cos \big[ \Phi   + \sum_{k=1}^N
( \alpha_k {\vec n}_k -i \xi_k {\vec v}_k ) \cdot {\hat {\vec \sigma }}_k
\big] +H.c. \big\} -\gamma_e \vec{S} \cdot \vec{H}_o +
{{\hat \tau }_z \over 2} \sum_{k=1}^N \omega_k^{\parallel} \:
 {{\vec l}_k \cdot {\hat {\vec \sigma }}_k } + \sum_{k=1}^N
\omega_k^{\perp}\: {{\vec m}_k \cdot {\hat {\vec \sigma }}_k } \;.
\label{4.1}
\end{equation}

There is also an inter-nuclear interaction of dipolar form; thus in general we
have to add to (\ref{4.1}) a term
\begin{equation}
\hat{V}( \{ \vec{\sigma}_k \} ) = \sum_{k\ne k^\prime } V_{k k^\prime }
^{\alpha \beta }
 \hat{\sigma}_k^\alpha   \hat{\sigma}_{k^\prime }^\beta \;.
\label{4.1o}
\end{equation}

We shall not need to write down the matrix elements $ V_{k k^\prime }
^{\alpha \beta }$ in detail; their strength is $\sim T_2^{-1}$, the transverse
nuclear relaxation rate.
Typically we shall assume that $\vec{H}_o$ is parallel to the easy axis of
our easy-axis/easy-plane system, so that the total effect of $\vec{H}_o$ will
be
written as
\begin{equation}
 \gamma_e \vec{S} \cdot \vec{H}_o \longrightarrow \xi_H \hat{\tau}_z \;,
\label{4.2}
\end{equation}
where $\xi_H = \gamma_e SH_o$
(we change notation - the symbol "$\epsilon$" will be used
to denote the "internal" bias coming from the nuclei).
The spin bath is described in (\ref{4.1}) by
spin-$1/2$ variables $\{ \vec{\sigma}_k \}$. The vectors ${\vec n}_k$,
${\vec v}_k$, ${\vec l}_k$, and  ${\vec m}_k$ are unit vectors; the
"static" couplings between $\vec{S}$ and the $\{ \vec{\sigma}_k \}$ are
described by the last two terms, and the dynamic effects are given by the
terms in the curly brackets. The longitudinal static coupling
$\omega_k^{\parallel}$ tells us the change in energy of $\vec{\sigma}_k$
when $\vec{S}$ flips between $\pm \hat{\vec{z}} S$; thus the
vector $\omega_k^{\parallel} \: {\vec l}_k$ represents the difference
between the effective field
acting on $\vec{\sigma}_k$ before $\vec{S}$ flips and that field after
$\vec{S}$ flips. If the $\{ \vec{\sigma}_k \}$ describe nuclei, this field will
be almost entirely hyperfine in origin, unless the $\{ \vec{\sigma}_k \}$
are produced by truncating the levels of a higher-spin nucleus (i.e., a
nuclear moment $\vec{I}_k$ with $\mid \vec{I}_k \mid  >1/2$).
In this latter case the total field acting on the $\{ \vec{\sigma}_k \}$ will
also include ,e.g., quadrupolar contributions; it then follows that the field
$\omega_k^{\perp}\: {\vec m}_k$ will be non-zero. This transverse field
is the {\it sum} of the fields acting on $\vec{\sigma}_k$ before and after
$\vec{S}$ flips.

The dynamic terms in (\ref{4.1}) originate as follows. The phase $\Phi =
\pi S +\sum_k \phi_k $ is the sum of the "Kramers phase" $\pi S$ (see
Introduction - notice that in (\ref{4.1}) we have not yet included
$\cos \pi S$ into the definition of $\Delta_o$) and a renormalisation
$\sum_k \phi_k $ of this phase, caused by the $\{ \vec{\sigma}_k \}$. The term
$\alpha_k$ describes the amplitude for $\vec{\sigma}_k$ to flip, under
the influence of $\vec{S}$, when $\vec{S}$ flips; both $\phi_k$ and $\alpha_k$
partly derive from the unitary "transfer matrix" $\hat{T}_k$ describing
the effect of a tunneling $\vec{S}$ on $\vec{\sigma}_k$:
\begin{equation}
 \mid \chi_{k}^{fin} \rangle  = e^{-i\int d\tau H_{int} (\tau ) }
\mid \chi_{k}^{in} \rangle \equiv \hat{T}_k^\pm
\mid \chi_{k}^{in} \rangle \;;
\label{4.3}
\end{equation}
\begin{equation}
 \hat{T}_k^\pm = e^{
\pm i (\alpha_k {\vec n}_k  \cdot {\hat {\vec \sigma }}_k
 +\phi_k  ) } \;.
\label{4.4}
\end{equation}
Here $H_{int} (\tau )$ describes the microscopic interaction between
$\vec{S}$ and $\vec{\sigma}_k$; $\mid \chi_{k}^{in} \rangle$
is the environment state before $\vec{S}$ flips, and
$\mid \chi_{k}^{fin} \rangle$ after $\vec{S}$ flips. The signs $\pm$ refer
to the path traced by $\vec{S}$ whilst flipping. The term in $\xi_k$
and further contributions to $\phi_k$ and $\alpha_k$ are necessary because
environmental spins also have an effect on the tunneling action for
$\vec{S}$ (in exact analogy with the nondiagonal terms discussed in Section
\ref{sec:3}).

The derivation of (\ref{4.1}) has been described in detail elsewhere \cite{10}.
Here we shall simply quote the values of the parameters in (\ref{4.1}) that
arise if we start with the microscopic hyperfine
Hamiltonian
\begin{eqnarray}
H_{Hyp}&=& H_o^{\perp} (\vec{S}) + {1 \over S}
\sum_{k=1}^N \omega_k \vec{S} \cdot {\vec I}_k \nonumber \\
&=&
{1 \over S}\bigg[ (-K_{\parallel}\:S_z^2 + K_{\perp}S_y^2)+
\sum_{k=1}^N \omega_k \vec{S} \cdot {\vec I}_k \bigg] \;,
\label{4.5}
\end{eqnarray}
i.e., our usual easy-axis/easy-plane model coupled to nuclei via
hyperfine couplings $\omega_k$; we get
\begin{equation}
\begin{array}{cclc}
\Phi & =& \pi S& \;\;\;\;(\phi_k=0)\\
\xi_k &=& \alpha_k/\sqrt{2}
 = \pi \omega_k/2\Omega_o&\;\;({\rm if}\; \omega_k \ll
\Omega_o) \\
\omega_k^{\parallel} &=& \omega_k & \\
\omega_k^{\perp} &=& 0&
\end{array}
\label{4.6}
\end{equation}
\begin{equation}
\begin{array}{ccl}
\vec{n}_k &=& (\hat{\vec{x}},\:\hat{\vec{y}})/\sqrt{2} \\
\vec{v}_k &=& -\hat{\vec{x}} \\
\vec{l}_k &=& \hat{\vec{z}}
\end{array}
\label{4.7}
\end{equation}

The assumption in (\ref{4.6}), that the hyperfine coupling
$\omega_k \ll \Omega_o$, is almost always true. In any case
$\alpha_k$, $\xi_k$, and $\phi_k$ can be calculated as general functions of
$\omega_k /\Omega_o$, for any initial microscopic  Hamiltonian. It is
useful to bear in mind the physical meaning of the parameter $\alpha_k$
 in this regime; $\alpha_k$ is the amplitude
for $\vec{\sigma}_k$ to coflip with $\vec{S}$. Note also that more
general interactions than (\ref{4.5}) will produce further renormalisations
of $\Delta_o$.

Working with (\ref{4.1}) (which is of course nothing but a low-energy
effective Hamiltonian, in the usual spirit which also led to (\ref{2.4}) and
(\ref{3.1})) allows us to bring out the 4 essential physical mechanisms
operating in the dynamics of $\vec{S}$. These are

(i) Even when $\vec{H}_o =0$, an internal field bias $\epsilon =
\sum_{k=1}^N \omega_k^{\parallel} \sigma_k^z$ acts on $\vec{S}$. Typically
for nuclei we expect the hyperfine couplings to be tightly clustered
around a principal value $\omega_o$, with $\omega_o \gg \Delta_o$.
There will however be a spread $\delta \omega_k $, due not only to
inter-nuclear couplings (principally via dipolar or
Nakamura-Suhl interactions, with
$ \delta \omega_k \sim 10^4-10^6\:Hz$) but also other couplings such as
transfer
hyperfine couplings.
One may then define a "density of states" $W(\epsilon )$ for bias
$\epsilon$ (Fig.3); if $\mu = N^{1/2} \delta \omega_k /\omega_o \ll 1$, this
consists of Gaussian  peaks
\begin{equation}
G_\mu (\epsilon - \omega_o \Delta \!N/2 )={1 \over \Gamma_\mu \pi^{1/2} }
\exp \{ -(\epsilon - \omega_o \Delta \!N/2 )/\Gamma_\mu ^2 \} \;,
\label{G}
\end{equation}
of width
\begin{equation}
 \Gamma_\mu = \mu \omega_o \;,
\label{GG}
\end{equation}
 around $\epsilon =
\omega_o \Delta \!N/2$, where $\Delta \!N = N_{\uparrow} - N_{\downarrow}$ is
the total nuclear polarisation, inside a Gaussian envelope of width
$N^{1/2} \omega_o$; this envelope extends out to a maximum bias
$\pm N\omega_o$:
\begin{equation}
W(\epsilon ) = 2^{-N} \sum_{\Delta \!N}
C^{(N+\Delta \!N )/2}_N G_\mu (\epsilon - \omega_o \Delta \!N/2 ) \;.
\label{4.8a}
\end{equation}
 (Here we assume that $N=$even integer; otherwise
$\Delta \!N =0$ is impossible! The modifications required for $N=$odd are
trivial and irrelevant to the ensuing discussion). In most cases, however,
$\mu >1$, and the different "polarisation groups" (i.e., the different peaks)
completely overlap, and we end up with a Gaussian envelope for $W(\epsilon )$,
i.e.,
\begin{equation}
W(\epsilon ) \longrightarrow {f \over \omega_o }
\exp \{ -2\epsilon^2/ (\omega_o^2 N) \}  \;; \;\;\;(\mu >1) \;.
\label{4.8}
\end{equation}
where $f=\sqrt{2/\pi N}$.

Now for tunneling to occur at all, the total bias
$(\epsilon +\xi_H)$ must be $\le \Delta_o$,
otherwise the grain is simply trapped. The fraction of grains in an ensemble
having bias $\le \Delta_o$ is roughly
$A \sim \Delta_o /N\delta \omega_k $ if $\mu \ll 1$, and $A \sim \Delta_o /
N^{1/2} \omega_o$ if $\mu > 1$; the result for $\mu \ll 1$ arises because
only a fraction $\sim \Delta_o / N^{1/2}\delta \omega_k $ of that
portion $f=\sqrt{2/\pi N}$ of grains having $\Delta \!N=0$ can flip when
the different polarisation groups do not overlap.

Thus only a small fraction $A$ of spins are not "degeneracy blocked"
from tunneling (in the absence of nuclear dynamics, the effects of which are
discussed below and in a Section \ref{sec:5}). This degeneracy blocking
mechanism operates just as easily whether there is an external bias or not - in
both cases only grains in the small window of bias can make transitions.

(ii) Even in the absence of an external thermostat, the nuclear system
can change the internal bias $\epsilon$. This occurs via
dipolar interactions between the nuclei, at a rate $T_2^{-1}$. This
pairwise flipping of nuclei conserves $\Delta \!N$ but allows
the nuclear bath to "wander in bias space" over the full range of
bias energy associated with a particular polarization group, i.e., over an
energy range $\sim \mu \omega_o = N^{1/2} \delta \omega_k$. If the spread
$ \delta \omega_k$ is due entirely to dipolar interactions, so
$\delta \omega_k \sim T_2^{-1}$, then the fluctuating bias covers the
whole energy range in a time $T_2$. This fluctuating bias $\epsilon (t)$
can destroy coherence, but it can help magnetic relaxation, by helping
the system find the small bias window.

(iii) The third mechanism, called "orthogonality blocking", arises because
the transverse fields $\omega_k^{\perp} $ acting on the environmental
spins (which are typically $\ll \omega_k^{\parallel}$) cause a mismatch
between the initial and final nuclear wave-functions, in a way reminiscent of
Anderson's catastrophe \cite{29}. Defining $\sin 2\beta_k =
\omega_k^{\perp}/ \omega_k^{\parallel} $ (assuming $\beta_k \ll 1$), this
blocking effect can be parametrised by $\kappa$, where
\begin{equation}
e^{-\kappa } = \prod_{k=1}^N \cos \beta_k  \;,
\label{4.9}
\end{equation}
so that $\kappa \approx 1/2 \sum_k \beta_k^2$. Orthogonality blocking
inhibits both coherence and relaxation of $\vec{S}$; it spreads out the
high-frequency ($\sim \tilde{\Delta}_o$) response of the grain to lower
frequencies.

(iv) A final interaction mechanism arises simply because the giant
spin can flip the nuclear spins. This process is parametrised by the
$\alpha_k$ in (\ref{4.4}); for the whole spin bath one finds that on
average a number $\lambda$ of spins will be flipped each time $\vec{S}$
flips, where $\lambda =  1/2 \sum_k \alpha_k^2$ for $\alpha_k \ll 1$.
This causes phase decoherence; we have called this mechanism
"topological decoherence" because it adds a random winding number to the
effective action for $\vec{S}$. It has a decoherence effect on tunneling,
and thus inhibits magnetic relaxation.

The formal discussion of all these mechanisms has appeared in our previous
work. In the next 2 sub-sections we will deal first with the dynamics of
$\vec{S}$ in the absence of spin diffusion between the nuclei, to show how
the nuclei keep all but a tiny fraction of grains frozen; then we
show how spin diffusion changes this picture.

\subsection{Tunneling Rate in a Bias}

We assume that neither the grain nor the spin bath is connected
to an external thermostat except at time $t \le 0$ (in reality
both are coupled to phonons and possibly electrons - see next section).
We also neglect nuclear spin diffusion here, and estimate its role later.
Thus relaxation of $\vec{S}$, in a bias, can only occur if the bias field
due to the $\{ \vec{\sigma}_k \}$ allows states $\mid \Uparrow ;
\{ \vec{\sigma}_k^{in} \} \rangle$, with the $\{ \vec{\sigma}_k^{in} \}$ in
some polarisation state $\Delta \!N^{in}$, to overlap in energy
(within $\sim \Delta_o$) with some states $\mid \Downarrow ;
\{ \vec{\sigma}_k^{fin} \} \rangle$, where the polarisation state
$\Delta \!N^{fin} \ne \Delta \!N^{in}$ in general. Thus we expect some nuclear
spins to flip, if $\vec{S}$ is to relax, although if $\mu $ is large, and
the applied bias is small one can even have energy overlap between
initial and final states having the same $\Delta \!N \ne 0$ (this does not, of
course, mean that no spins are flipped during the transition, but
only that none have to be flipped).

The formal treatment of this problem is a generalisation of
our treatment of the unbiased
case. We write, for the time correlation function
$P(t) = \langle \hat{\tau}_z(t)
\hat{\tau}_z(0) \rangle $, the form \cite{10}
\begin{equation}
 P(t;T;\xi_H ) =1 + \int d\epsilon {W(\epsilon )e^{-\beta \epsilon }
\over Z(\beta )}
 \sum_{M=-N/2}^{N/2} \big[
P_M(t,\epsilon +\xi_H -M\omega_o/2 )-1 \big] \;.
\label{4.10}
\end{equation}

This formula is crucial, and so we now spend a little time explaining it.
The average $\int d\epsilon$ is over initial bias, with both thermal and
density of states weighting. $P_M(t,\epsilon )$ is a function
defined in Ref.\cite{10}; it describes all those grains in an ensemble
having bias $\epsilon$ and for which, every time $\vec{S}$ flips,
the polarisation state of the $\{ \sigma_k \}$ changes by $2M$:
\begin{equation}
 P_M(t) = \int_0^\infty dye^{-y} \sum_{\nu=-\infty}^{\infty} \int {d\varphi
\over 2 \pi } F_{\lambda^\prime }(\nu )
e^{2i\nu (\Phi -\varphi )} P_M^{(0)} (t,\epsilon ,\varphi ,y)  \;,
\label{4.11}
\end{equation}
in which $ P_M^{(0)}$ describes a simple biased 2-level system
(cf. Eq.(\ref{1.7}))
\begin{equation}
 P_M^{(0)} =1-  {\Delta_M^2(\varphi ,y) \over E_M^2 (\varphi ,y) }
\sin ^2 (E_M (\varphi ,y)t) \;,
\label{4.12}
\end{equation}
\begin{equation}
E_M^2 = \epsilon^2 + \Delta_M^2 \;,
\label{4.13}
\end{equation}
\begin{equation}
\Delta_M(\varphi ,y)=  2{\tilde \Delta}_o \mid \cos \varphi J_M
(2\sqrt{(\lambda -\lambda^\prime)  y}) \mid \;,
\label{4.14}
\end{equation}
with  $\lambda = 1/2 \sum_k \alpha_k^2$,
$\lambda^\prime  = 1/2 \sum_k \alpha_k^2 (n_k^z)^2$
(so that $\lambda \ge \lambda^\prime$).
Eq.(\ref{4.11}) for $ P_M(t)$ can be understood as combining a "phase average"
\begin{equation}
\sum_{\nu} \int {d\varphi
\over 2 \pi } F_{\lambda^\prime }(\nu )
e^{2i\nu (\Phi -\varphi )} \;; \;\;\;\;F_{\lambda^\prime }(\nu )=e^{-4\lambda
\nu^2} \;,
\label{4.14b}
\end{equation}
over the phase $\varphi $ and winding number $\nu $), with an "orthogonality
average"
\begin{equation}
\int_0^\infty dye^{-y}
\label{4.14c}
\end{equation}
These averages are performed over the biased 2-level correlation function
$P_M^{(0)} (t,\epsilon ,\varphi ,y)$, in which the tunneling
amplitude $\Delta_M(\varphi ,y)$ depends on $M$, $\varphi$, and $y$
via (\ref{4.14}).

Eq.(\ref{4.10}) for
 $P(t)$ may seem a little odd, since it sums over apparently
independent processes in which the polarisation state of the nuclei
changes by $2M$ each time $\vec{S}$ flips. Why should the polarisation
state change by $2M$ each time? After all, when $\vec{S}$ flips,
the number of nuclei which flip is {\it random} (with an average number
$\lambda $)!
 The answer to this question
comes from  energy conservation considerations.

Consider a single grain that at $t=0$ has its nuclear environment in the state
$\mid \chi_1,\epsilon_1, \Delta \!N \rangle $ corresponding  to
 the polarisation group $\Delta \!N$ and the  bias energy
$\epsilon_1$ (see Fig.4).
Now $\vec{S}$ flips - what are the possible final states for the
combined system, which have energies in resonance with $\epsilon_1$?
There are three possibilities:

(1) None of the environmental spins flip; then the final state is
    $\mid \chi_1, \epsilon_f, -\Delta \!N )\rangle
    $ with $\epsilon_f = -\epsilon_1$,
    since $\tau_z \sigma_k^z$ changes sign.

(2) Some small number of environmental spins, say $r \ll N$, flip
    together with $\vec{S}$, so that the net change in the nuclear spin
    polarisation is $2M$ (i.e.,  $(r+M)/2$ up and $(r-M)/2$ down spins
    are flipped).  As a first approximation the final
    state    energy will be
    $\epsilon_f \approx -\epsilon_1 +M\omega_o $; assuming that the spread in
    nuclear frequencies is small, we may neglect for the moment the
    correction $\delta \epsilon_r \sim \delta \omega_k \sqrt{r}
    \ll (\mu \omega_o, \omega_o) $. We denote these states as
    $\mid 2 \rangle $, and $\mid 3 \rangle $ in Fig.4.

(3) An enormous number of nuclear spins flip together with $\vec{S}$,
    with $r$ as large as $N$; in this case we can not neglect the
    correction $\delta \epsilon_{r\sim N } \sim \mu \omega $ any more, because
    it is comparable or even larger (for $\mu >1$)  than the typical
    difference $Min_{\{ M \} } (2 \epsilon -M\omega_o ) \sim \omega_o $.
    Then we can use $\delta \epsilon_{r}$ to
   "fine  tune"
    a  resonance ( which is impossible, in general, for the case (2)).
    Thus, we find in this case $ \epsilon_f = -\epsilon_1 +M\omega_o +
     \delta \epsilon_{r}$. The correction strongly depends, of course,
    on the particular set of nuclei flipped; a "fine tuned" situation is shown
     as state $\mid 4 \rangle $ in Fig.4.

It seems at first, that the latter possibility is the best we can do
for tunneling in resonance.
Recall, however, that the probability that any  given environmental
spin flips during the transition is very small ($\alpha_k^2 \ll 1$).
Now, if we
are going to tunnel between $\mid 1 \rangle $ and $\mid 4 \rangle $ then
the amplitude of such a transition will be $\sim  \Delta_o (\alpha_k^2)^N $,
 and for large $N$ will be even less
then $ 2^{-N} \delta \omega_k$! Thus such
a resonance is simply impossible to realise.

In fact, only a small number of environmental spins may be flipped with a
reasonable probability. Since the probability to flip none is just
$e^{-\lambda}$, we find the probability to flip exactly $r$ spins to be
$p_r \approx e^{-\lambda} (\alpha_k^2)^rC^N_r \approx e^{-\lambda}
\lambda^r/r!$,
which peaks at $r \sim \lambda$. The parameter $\lambda$ may be large
in some systems , but still $\lambda \ll N$. One immediately recognizes that
the case (1) plays a role at  small $\lambda$,  while  the case (2) will
dominate when $\lambda \gg 1$ (formally, we could include (1) in (2) as a
particular transition with $r=0$).

Now we make use of the inequality $\omega_o \gg \Delta_o$ to notice that
among all possible transitions with different $M$, the energy mismatch between
the initial and final states $ \epsilon_1-\epsilon_f = 2\epsilon_1 -M\omega_o$
is either much larger then $\Delta_o$ for all $M$, or is close to
the resonance $\epsilon_1-\epsilon_f \sim \Delta_o$ for only one specific
value $M_{\epsilon_1}$, with all the other transitions being $\ge \omega_o$
away in energy. Once we have a resonance the system will make transitions
from $\Uparrow $ to $\Downarrow$ by changing the nuclear spin polarisation
by exactly $\pm 2 M_{\epsilon_1}$ each time to maintain this resonance.
As for all the other transitions with $M \ne M_{\epsilon_1}$,
they give only very small corrections, of order
$(\Delta_o /\omega_o )^2$ or less (cf. Eqs.(\ref{1.7}) and
(\ref{4.12})) to the main contribution to the dynamics
of $\vec{S}$, coming from the $M=M_{\epsilon_1}$ term. Of course
Eqs.(\ref{4.10}) and (\ref{4.11}) simply sum over all processes,
and the resonance value of $M$ is included automatically

It is worth noting that nothing depend explicitly on the initial
polarisation state. This point is further illustrated in Fig.4.
For given $\epsilon_1$ the initial polarisation state can be $M$, and if
 $\epsilon_1$ is close to the center of the Gaussian $G_\mu (\epsilon -
M\omega_o/2)$, then the resonance value is $M_{\epsilon_1}=M$. If however
$\epsilon_1$ is, say, the down-tail state of another polarisation group
$M+2$, then transition to the up-tail state of $-M+2$ has exactly the same
energy mismatch and polarisation change $2M$. That is why the {\it only
relevant statistical average is over the bias energy}. Of course,
if $\mu \ll 1$, and different polarisation groups do not overlap, the
bias energy is related to the specific group and Eq.(\ref{4.10}) may be
written as ($\xi_H =0$)
\begin{equation}
 P(t) ={1 \over 2^N Z(\beta ) }
\sum_M C^{(N+M)/2}_N
 \int d\epsilon G_\mu (\epsilon - M\omega_o/2 )e^{-\beta \epsilon }
P_M(t,\epsilon +\xi_H -M\omega_o/2 ) \;.
\label{4.19}
\end{equation}

It is easy to understand now that external bias, which is indistinguishable
from the internal one as far as the giant spin is concerned, will simply
shift the resonance condition for given $\epsilon$ to some other value
$M_{\epsilon+\xi_H}$, as is clearly seen from the diagram in Fig.5, which is
similar to Fig.4, but now with $\xi_H \ne 0$.

We may further simplify Eq.(\ref{4.10}) by changing variables from
$\epsilon +\xi_H -M\omega_o/2 \to \epsilon$, and write it as
\begin{equation}
 P(t;T;\xi_H ) =1 + \sum_{M} \int d\epsilon {W(\epsilon-\xi_H+M\omega_o/2  )
e^{-\beta (\epsilon-\xi_H+M\omega_o/2 ) }
\over Z(\beta )}
 \big[
P_M(t,\epsilon )-1 \big] \;.
\label{4.20}
\end{equation}
Then because the function $\big[
P_M(t,\epsilon )-1 \big] \sim (\Delta_o /\epsilon )^2 $ for large bias,
and in most cases $W(\epsilon )$ is a smooth function on a $\Delta_o$
scale (for $\mu \omega_o \gg \Delta_o$), we have
\begin{equation}
 P(t;T;\xi_H ) = \sum_{M}  {W(M\omega_o/2-\xi_H  )
e^{-\beta (M\omega_o/2-\xi_H ) }
\over Z(\beta )}
\int d\epsilon  P_M(t,\epsilon )\;, \;\;\;(\mu \omega_o \gg \Delta_o) \;.
\label{4.21}
\end{equation}
(we assumed here that $\beta \Delta_o \ll 1$). In the opposite limiting case
$\mu =0$, when the Gaussian in (\ref{4.19}) is a $\delta$-function, we have
\begin{equation}
 P(t) ={1 \over 2^N Z(\beta ) }
\sum_M C^{(N+M)/2}_N
e^{-\beta M\omega_o/2 }
P_M(t,0 ) \;.
\label{4.22}
\end{equation}

To summarize, we may understand (\ref{4.10}) in a fairly simple way as
including all effects of the spin bath on the dynamics of $\vec{S}$. These
come from the averages  over phase (Eq.(\ref{4.14b}), over the orthogonality
mismatch between initial and final spin bath states (Eq.(\ref{4.14c})), and
over the internal bias $\epsilon$, acting on $\vec{S}$, caused by the spin
bath. It also comes from summing over all possible changes in the polarisation
state of the bath when $M$ flips. It is perhaps worth noting, for those who may
be used to the theory of "oscillator bath" environments \cite{17}, that the
averages appearing here for the spin bath environment are very different
in form, for the simple reason that in the case of the spin bath, {\it most
of the dynamics of the spins comes from their coupling to the macroscopic
system itself} (in our case, to $\vec{S}$). By contrast in the oscillator bath
models, the coupling to the macroscopic system is weak ($\sim O(N^{-1/2})$, for
each oscillator), and the dynamics of the individual oscillators  is only
weakly perturbed by the system. This is why, in the theory of the oscillator
bath, one can first calculate the weakly perturbed oscillator motion
as a function of the system coordinates, and then integrate out the oscillators
by functional averaging. No such manoeuvre is possible for the spin bath
(or indeed any other environment where the couplings are not weak), and it
is {\it not possible} to map the spin bath onto an oscillator bath
(see also Ref.\cite{28}). Any attempt to do so (by, e.g., writing
a spectral function $J(\omega ) \sim N \delta (\omega -\omega_o)$, for a set
of "nuclear spin oscillators", as in ref.\cite{18}), cannot be expected
to give meaningful results.

{}From these remarks we can also see what has been left out of Eq.(\ref{4.11})
for $P_M(t)$; we have left out everything coming from the independent dynamics
of the spin bath (independent, that is, from $\vec{S}$). Formally this can be
put in by including these dynamics in $P_M^{(0)}$ in (\ref{4.12}).
As we argue below this corresponds to allowing a
{\it time-dependent bias} $\epsilon (t)$ in (\ref{4.10}) and (\ref{4.11}).

We neglect the effect of coupling of the nuclei to phonons which is actually
an incredibly small effect in most cases.
In reality the nuclear spin dynamics comes almost entirely from nuclear spin
diffusion, parametrised by the transverse relaxation time $T_2$, and caused
by the dipolar interaction in (\ref{4.2}). As we shall see presently, if the
spin bath is mainly composed of nuclei, this spin diffusion is crucial,
since without it $\vec{S}$ cannot relax at all. Spin diffusion processes,
in which $\Delta \!N$ remains unchanged but pairs of nuclei flip, cause
$\epsilon (t)$ to fluctuate in time because of the variation
$\delta \omega_k$ in the coupling of each nuclei to $\vec{S}$ (thus, for
a process $\mid \uparrow \downarrow \rangle \to \mid \downarrow \uparrow
 \rangle$ involving 2 nuclei, the total change in internal bias $\sim
\delta \omega_k$; if $N$ nuclei flip, the change in $\epsilon $ is
$\sim \delta \omega_k N^{1/2}$). Now this fluctuation is {\it fast},
in the sense that for large values of $M$ the time it takes
$\epsilon (t)$ to change by an energy $\Delta_M$  is usually
much less than the time $\Delta_M^{-1}$ required for $\vec{S}$ to flip
if it is in the coherence window (of energy width $\Delta_M$); this will
be demonstrated below. Consequently it is not necessary (or even useful)
to go through the elaborate detour of recalculating $P_M^{(0)}$ in
Eq.(\ref{4.12}), including a coupling to some effective "oscillator bath",
intended to model the effects of spin diffusion. We simply write instead
that for this fast diffusion problem, where $\epsilon (t)$ varies over an
energy range $\mu \omega_o$, one has:
\begin{equation}
 P(t;T;\xi_H )- {1 \over 2} = \sum_{M} { C^{(N+M-M_H)/2}_N
e^{-\beta (M-M_H)\omega_o/2 }
\over 2^NZ(\beta )} \bigg[ \langle P_M(t, \epsilon (\tau )+\delta \xi_H)
\rangle_{\epsilon (\tau )} - {1 \over 2} \bigg] \;,
\label{4.23}
\end{equation}
where we introduced the notation
$\xi_H \equiv M_H\omega_o/2 +\delta \xi_H $ with
$M_H$=integer and
$\mid \delta \xi_H \mid \le \omega_o/2$ to define the shift in the polarisation
change enforced by applying the external bias. Here (as before)
\begin{equation}
\langle P_M(t, \epsilon (\tau )+\delta \xi_H)
\rangle_{\epsilon (\tau )} =
 \int_0^\infty dye^{-y} \sum_{\nu=-\infty}^{\infty} \int {d\varphi
\over 2 \pi } F_{\lambda^\prime }(\nu )
e^{2i\nu (\Phi -\varphi )} \langle
P_M^{(0)} (t,\epsilon (\tau ) +\delta \xi_H ,\varphi ,y)
\rangle_{\epsilon (\tau )}  \;,
\label{4.24}
\end{equation}
\begin{equation}
\langle  P_M^{(0)} (t,\epsilon (\tau ) +\delta \xi_H ,\varphi ,y)
\rangle_{\epsilon (\tau )} -{1 \over 2}  = {1 \over 2}
e^{-t/\tau_M( \delta \xi_H)} \;,
\label{4.25}
\end{equation}
\begin{equation}
\tau_M^{-1}( \delta \xi_H) = 2\Delta_M^2 G_\mu (\delta \xi_H)  \;,
\label{4.26}
\end{equation}
and $\Delta_M(\varphi ,y)$ is given by (\ref{4.14}). In the usual case where
$\mu >1$, or if  the external bias is close to a multiple of $\omega_o$,
the expression for the effective relaxation rate simplifies to
\begin{equation}
\tau_M^{-1} = {2\Delta_M^2 \over \Gamma_\mu \pi^{1/2} }  \;.
\label{4.27}
\end{equation}

This concludes our formal discussion of the problem of quantum
relaxation of $\vec{S}$ when it is coupled to a spin bath. We now
use the results to find $P(t,\xi_H)$ for a few interesting cases.

We start by ignoring spin diffusion, in order to demonstrate the way
in which relaxation is blocked in its absence. Let us first consider
a really pathological case, in which there is no degeneracy blocking
at all, i.e., $\mu=0$. Then the function $W(\epsilon )$ simply becomes
a set of sharp lines, and {\it all states in a grain ensemble
may resonantly tunnel}, if $\delta \xi_H =0$.
If $\delta \xi_H \ne 0$, then no resonance is possible.
Initially there will be fast relaxation, involving processes where
$M$ is not large. At longer times the higher-$M$ processes take over -
recall that for large $M$, the transition amplitude $\Delta_M$ is
very small, since $\Delta_M \sim \Delta_o \lambda^{M/2}/M!$, which
collapses when $M \gg \sqrt{\lambda }$. To get some idea of the resulting
relaxation, consider what happens if $\xi_H =0$, and let us ignore
topological decoherence for simplicity (which makes no difference
for the long-$t$ asymptotics of the result). Then we have
\begin{eqnarray}
 P_M(t) &=& \int_0^\infty dye^{-y} P_M^{(0)} (t,y)  \nonumber \\
 &=& {1 \over 2} \int_0^\infty dy \big[1+\cos (2\Delta_M(y)t ) \big]\;,
\label{4.29}
\end{eqnarray}
where $\Delta_M(y) = \tilde{\Delta}_o J_M(2\sqrt{\lambda y })$,
assuming $\lambda^\prime =0$ (note if $\lambda^\prime =0$, then the
topological phase average (\ref{4.14b}) collapses to a $\delta$-function
$\delta (\varphi -\Phi )$ and we get (\ref{4.29}) anyway). The resulting
curves are shown for $\lambda =5$ in Fig.6, for various $P_M(t)$.
It is easy to then find the behaviour of $P(t)$ by substituting (\ref{4.29})
into  (\ref{4.22}). The steepest-descent integral over $M$ gives
an accurate answer for long times. However we more quickly
derive this long time behaviour from dimensional arguments. The sum over
$M$ contains some $\sqrt{2N}$ polarisation groups inside the broad
envelope function $C^{(N+M)/2}_N$; thus the crossover from $P(t) \approx 1$
to the equilibrium $P(t) \approx 1/2$ occurs around a time $t_c$ such that
$\Delta_{\sqrt{2N}} t_c \sim 1$, i.e.,
\begin{equation}
\ln (\tilde{\Delta}_o t_c) \sim \sqrt{ {N \over 2}} \ln \left(
{2N \over \lambda e^2 } \right) \;.
\label{4.30}
\end{equation}

For shorter times the relaxation is roughly logarithmic, viz.
\begin{equation}
1-P(t) \sim  \sqrt{ {2 \over N} }
 { \ln (\tilde{\Delta}_o t) \over
 \ln \big[ { \ln (\tilde{\Delta}_o t) \over e\sqrt{\lambda } } \big]}
\;; \;\;\;\;(t\ll t_c)  \;,
\label{4.31}
\end{equation}
whereas for longer times one has
\begin{equation}
P(t)-1/2 \sim \exp \bigg\{
-3 { \ln ^{2/3}(\tilde{\Delta}_o t) \over N^{1/3} } \bigg\} \;; \;\;\;\;
(t \gg t_c) \;,
\label{4.32}
\end{equation}
with logarithmic accuracy. The numerical evaluation of $P(t)$ is shown
in Fig.7.

However even the tiniest degeneracy blocking will upset these results - in
fact if $\delta \omega_k \sqrt{N} > \Delta_M$, the result (\ref{4.29})
fails completely.
Since the proverbial "thunderstorm on Jupiter" is enough to give a
$\delta \omega_k $ exceeding $\Delta_M /\sqrt{N} $ for the large values of $M$
governing long time relaxation, we see that for relaxation (just as for
coherence \cite{10}), it is the limit of strong degeneracy blocking that
is experimentally meaningful. Let us again consider the zero bias
case in this limit, where $\mu > \Delta_o /\omega_o$. We go back
to (\ref{4.21}), again assume $\lambda^\prime =0$ for simplicity, and,
noting that the oscillatory $\sin^2 E_Mt$ term in $P_M^{(0)}$
(Eq.(\ref{4.12})) gives an integrated contribution
\begin{equation}
\int d\epsilon {\Delta_M^2 (y) \over \epsilon^2 +\Delta_M^2} \sin ^2
[t\sqrt{\epsilon^2+\Delta_M^2(y)} ] = {\pi \Delta_M(y) \over 2}
\int_0^{2\Delta_M(y)t} dz J_0(z) \;;
\label{4.33}
\end{equation}
we find
\begin{equation}
1-P(t)=\sum_M {W(M\omega_o/2) e^{-\beta M\omega_o/2} \over Z(\beta )}
\int_0^\infty dy   {\pi \Delta_M(y) \over 2}
\int_0^{2\Delta_M(y)t} dz J_0(z) \;;
\label{4.34}
\end{equation}
This result describes an ensemble of grains in which a fraction
$A$ of grains relaxes, leaving a fraction $1-A$ completely unrelaxed in the
infinite-time limit; A is given by
\begin{equation}
A=1-P(t \to \infty )= {\pi \tilde{\Delta}_o \over 2} \sum_M {W(M\omega_o/2)
 e^{-\beta M\omega_o/2} \over Z(\beta )}
\int_0^\infty dy \mid J_M(2 \sqrt{\lambda y }) \mid   \;;
\label{4.35}
\end{equation}
Since terms for $M > \sqrt{\lambda }$ make almost no contribution to
$A$ due to the  collapse of $\Delta_M$, it can be approximated for large
$\lambda $ by
\begin{equation}
A=\lambda^{1/4} {\tilde{\Delta_o} \over \omega_o  }f
e^{-N\omega_o^2 /8(k_BT)^2 }   \;,
\label{4.36}
\end{equation}
where $f=\sqrt{2/\pi N}$ is the number of states in the $\Delta \! N =0$
polarisation group (in Ref.\cite{10} we showed that individual terms in the sum
over $M$ decreased as $\lambda^{-1/4}$; but the sum over all $M$ gives the
factor $\lambda^{1/2}$).

Thus we see that apart from the small number of grains in near resonance,
most grains will be frozen for eternity by the nuclear bias field, even in
the absence of an external bias (again, ignoring nuclear spin diffusion).
Adding an external bias makes essentially no difference to this
(since this bias is physically indistinguishable from the internal bias).
In fact the bias will have no effect at all on (\ref{4.34}) until
$\xi_H \sim $ the energy scale of variation of $W(\epsilon )$.
For $\mu <1$, this means that we will see oscillations in the decay rate
as a function of $\xi_H$, with period $\omega_o$. For $\mu >1$, the decay rate
and decay function $A$ will change very little until $\xi_H \sim \omega_o
\sqrt{N}$; for $\xi_H \gg \omega_o
\sqrt{N}$  the relaxed fraction $A \to 0$ since not even the nuclei
can bring the system into resonance. We derive also a useful formula
for the bias and temperature effects by noting that only small
$M \ll \sqrt{N}$ contribute to the sum in Eq.(\ref{4.21}) and
thus $W(M\omega_o/2 -\xi_H)\exp \{ -\beta (M\omega_o/2-\xi_H) \}
\approx W(-\xi_H)\exp \{ \beta \xi_H) \}$ giving
\begin{equation}
P(t;T;\xi_H) =P(t;\infty ;0) {W(\xi_H)e^{ \beta \xi_H }
\over W(0)Z(\beta ) }
\; .
\label{4.36f}
\end{equation}

None of the results (\ref{4.29})-(\ref{4.36f}) is physical at very long time,
because we have ignored the time variation of $\epsilon $ caused by nuclear
spin diffusion. Let us now demonstrate that nuclear spin diffusion is
{\it fast}, so that equations (\ref{4.24})-(\ref{4.26}) are justified.
Notice first that since dipolar pairwise flips occur at a rate $T_2^{-1}$
for a given pair, and assuming short-ranged pair flips, we have roughly
$NT_2^{-1}$ nuclear flips per second  in the sample. Then in a time
$\Delta_M^{-1}$ we have $N(T_2 \Delta_M)^{-1}$ such flips, and in this
time the bias will change by $\delta \epsilon \sim \delta \omega_o
\sqrt{NT_2 /\Delta_M}$. The condition for fast diffusion is
$\delta \epsilon \gg \Delta_M$ (so the system has no time to tunnel), i.e.,
fast diffusion requires \cite{5}
\begin{equation}
\Delta_M^3 \ll {N \over T_2 } (\delta \omega_o )^2
\; .
\label{4.37}
\end{equation}

 If the spread $\delta \omega_o $ arises from the dipolar interactions
between the same kind of nuclei only, then $\delta \omega_o \sim T_2^{-1}$
(but usually $\delta \omega_o > T_2^{-1}$, because of other nuclei and fields),
then this criterion becomes $\Delta_M \ll T_2^{-1} N^{1/3}$. Now in fact
$T_2^{-1}$ will be typical $10^3-10^6\:Hz$ (depending on the isotopic
concentration of nuclear spins, etc.). For any mesoscopic spin
$\tilde{\Delta}_o <10^6\:Hz$ (at least for most bare Hamiltonians), so that
even if (\ref{4.37}) is violated for $M=0$ (i.e., the zero polarisation
change process), it is  obeyed for large values of $M$. Since the bulk of the
relaxation involves values of $M \sim N^{1/2}$, we see that
(\ref{4.37}) will always be the relevant condition. In this case
Eqs.(\ref{4.23})-(\ref{4.27}) follow immediately.

Let us now therefore evaluate $P(t)$ for the coupled grain/spin bath
system, taking full account of the dipolar interactions between the spins,
and assuming (\ref{4.37}) to be satisfied; using
(\ref{4.23})-(\ref{4.25}), with $\lambda^\prime$, this means we must
evaluate, for $\mu > \Delta_o/\omega_o$
\begin{equation}
 P(t)- {1 \over 2} = {1 \over 2} \sum_{M} C^{(N+M-M_H)/2}_N
{e^{-\beta \omega_o(M-M_H)/2 }
\over 2^NZ(\beta )} \int_0^\infty dye^{-y}
\exp \bigg\{ -2 {\tilde{\Delta}_o^2 \over \pi^{1/2} \Gamma_\mu }
tJ_M^2(2\sqrt{\lambda y}) \bigg\} \;.
\label{4.38}
\end{equation}
The integral is very easily evaluated; calling it $I$, we have
\begin{eqnarray}
I & \sim &  \int_0^\infty dye^{-y}
\exp \bigg\{ -2 {\tilde{\Delta}_o^2t \over \pi^{1/2} \Gamma_\mu }
\left( {\lambda e^2 \over M^2 } \right) ^M y^M \bigg\} \nonumber \\
& \sim &  \int_0^\infty dye^{-y}
\theta (y_M(t) -y )  \;,
\label{4.39}
\end{eqnarray}
where the step-function $\theta (y_M(t) -y )$, with
\begin{equation}
y_M(t) = {M^2  \over \lambda e^2 } \big( \pi^{1/2}
\Gamma_\mu /2 \tilde{\Delta}_o^2t \big) ^{1/M} \;,
\label{4.40}
\end{equation}
arises because values of $M$ range up to roughly $\sqrt{2N}$, and are
thus typically large; thus
\begin{equation}
 P(t)-{1 \over 2} \approx {1 \over 2}  \sum_{M} C^{(N+M-M_H)/2}_N
{e^{-\beta \omega_o(M-M_H)/2 }
\over 2^NZ(\beta )}   \bigg\{ 1- \exp \bigg[
-{M^2  \over \lambda e^2 } \big( \pi^{1/2}
\Gamma_\mu /2 \tilde{\Delta}_o^2t \big) ^{1/M}
 \bigg] \bigg\} \;.
\label{4.41}
\end{equation}

Noting again that if $\mu >1$, the dependence of $P(t)$ on $\xi_H$ in
(\ref{4.41}) will be rather small until $\xi_H \sim \omega_o \sqrt{N}$,
we begin by analysing this result for $\xi_H =0$. This zero bias case
has of course been previously analyzed in our coherence papers \cite{4,5},
but there we were only interested in the very short time behaviour
of $P(t)$, i.e., that part involving frequencies $\sim \tilde{\Delta}_o$,
connected with possible coherent oscillations. Coherent behaviour can only
arise in the $M=0$ polarisation sector of (\ref{4.29}), which ignores
spin diffusion.

However in analysing the quantum relaxation properties of $\vec{S}$,
we need to sum over all $\sqrt{2N}$ important terms in (\ref{4.41}); the
effect of the spin diffusion will be to unblock the long-time relaxation.
In evaluating (\ref{4.38}) or (\ref{4.41}), a steepest-descent integration
over $M$ is possible, but just as with the case of no spin diffusion,
dimensional analysis is sufficient; the crossover time from unrelaxed to
almost relaxed behaviour occurs at a time $t_c$ given by
\begin{equation}
\ln \left( 2{\tilde{\Delta}_o^2 \over \pi^{1/2} \Gamma_\mu } t_c \right) \sim
\sqrt{2N} \ln \left(
{2N \over \lambda e^2 } \right) \;,
\label{4.42}
\end{equation}
(compare (\ref{4.30})), and the short-time relaxation is roughly logarithmic,
as in (\ref{4.31}), viz.,
\begin{equation}
1-P(t,\xi_H=0) \sim  \sqrt{ {1 \over2\pi  N} }
 { \ln \left( 2\tilde{\Delta}_o^2t/\pi^{1/2} \Gamma_\mu  \right) \over
 \ln \bigg[ {1 \over e\sqrt{\lambda } } \ln
\left( 2\tilde{\Delta}_o^2t/ \pi^{1/2} \Gamma_\mu \right)  \bigg] }
\;; \;\;\;\;(t\ll t_c)  \;,
\label{4.43}
\end{equation}
The  long-time behaviour is, analogously to (\ref{4.32}), given by
\begin{equation}
P(t,\xi_H=0) -1/2 \sim \exp \bigg\{
-{3 \over (2N)^{1/3}} \ln ^{2/3}
\left( 2\tilde{\Delta}_o^2t /\pi^{1/2} \Gamma_\mu \right)
 \bigg\} \;; \;\;\;\;
(t \gg t_c) \;,
\label{4.44}
\end{equation}
with $\ln \ln $ corrections in the exponent. Thus, amusingly, the results
with spin diffusion included look just like those {\it without} spin
diffusion, provided $\mu =0$ in the latter. The physical reason for this
is simple - rapid spin diffusion basically eliminates the effects of
degeneracy blocking, by allowing the bias $\epsilon (t)$ to cover the whole
energy range of each polarisation group. This is then the fundamental reason
why spin diffusion "unlocks" $\vec{S}$ and allows it to relax.

In Fig.7 we showed a plot of $P(t)$ against $\ln (2 \tilde{\Delta}_ot)$,
which clearly brings out the 3 relaxation regimes. For very short times
$\sim \tilde{\Delta}_o^{-1}$, there is a sudden relaxation involving
only $M \sim \lambda^{1/2}$ processes - the theory of this was given entirely
in our coherence papers. For times $t$ such that $\tilde{\Delta}_o^{-1}
\ll t \ll t_c$, we have roughly logarithmic relaxation. Then for
$t \gg t_c$, we have the behaviour in (\ref{4.32}) or (\ref{4.44}).

At this point it is important to realise that for all but microscopic
spins, $t_c$ will be astronomically long. From (\ref{4.42}) we have
\begin{equation}
t_c \sim {\pi^{1/2} \Gamma_\mu  \over 2\tilde{\Delta}_o^2 }
 \left( {2N  \over \lambda e^2 }  \right)^{\sqrt{2N}} \;.
\label{4.45}
\end{equation}

Consider now, as in Fig.7, the case of a mesoscopic spin where, e.g.,
$\tilde{\Delta}_o \sim 1\:MHz$, $\Gamma_\mu =  100\:MHz$, $N=1000$, and
$\lambda =10$. We then find that $t_c \sim 10^{50}\: s$, i.e., more
than $10^{30}$ times the age of the universe! Thus for mesoscopic
grains we {\it always} get logarithmic relaxation at long times,
from nuclear spins, in the quantum regime. Readers familiar with the
experimental measurements
of magnetic relaxation in grains may well be quite surprised at this result,
since it is almost a central dogma of magnetism that logarithmic relaxation
must arise from a distribution of energy barriers \cite{10}. Nevertheless
we see that not only is this not so, but that the logarithmic relaxation
has been derived here in a very general way, with essentially no
approximations (but see below).

On the other hand for microscopic spins ($S \sim 10$), one finds that
$t_c$ can be short. For example, consider a situation where
$S=20$, $\Omega_o =10^{10}\:GHz$,
$\tilde{\Delta}_o \sim 100\:MHz$, $\Gamma_\mu 100\:MHz$, $N=10$,
and $\lambda =0.1$ (this would roughly describe a particle containing
$10$ Tb atoms). Then one finds $t_c \sim 1\:s$, and any direct relaxation
experiment (typically conducted over time periods $1\:ms < t < 10^4\:s$)
would be mostly in the long-time relaxation regime.

These results are both surprising and interesting, in view of the very
long time (often logarithmic) relaxation observed in experiments. However
we immediately point out that the results (\ref{4.43})-(\ref{4.45})
are by no means complete - although they are correct to within
$\ln \ln $ corrections for a coupled grain/spin  bath system, a realistic
calculation must incorporate {\it all three} of the relaxation mechanisms
we have discussed, in Sections \ref{sec:2}, \ref{sec:3}, and this one.
Since this changes the results yet again, we will not waste time here
analysing the case of large bias ($\xi_H > \omega_o \sqrt{N}$ in
(\ref{4.23})-(\ref{4.26})) for the grain/spin bath system, but
proceed directly to the general case.

\section{Combining the Mechanisms - Physical Results}
\label{sec:5}

We now come to the crux of the paper, which involves putting together
the various mechanisms discussed in Sections \ref{sec:2}-\ref{sec:4} to give
a physically  realistic picture of quantum relaxation in magnetic
particles. As mentioned in the introduction, we do not attempt a
comprehensive discussion of all possible cases. Such a discussion  would
involve consideration  of a large class of "giant spin" bare Hamiltonians,
with widely varying behaviour, and with each being considered throughout the
whole range of field strength, field orientation, and temperature. This would
only be the beginning - we would then have to go on to examine the large
variety
of couplings to nuclear spins (with hyperfine couplings ranging over 3 orders
of magnitude, plus quadrupolar couplings), the variety of different
magnetoacoustic
couplings, etc., etc.

We adopt the tactic here of showing how the mechanisms combine, and giving
some details for one model. In other papers, in preparation, we analyse
some specific experimental systems in much greater detail. We emphasize
once again here that, in our view, magnetic relaxation in most
experiments involves the  dipolar interactions {\it between} grains
in an essential way.

We begin by showing how phonons and nuclear spins combine to give
a rather surprising form for the relaxation at low $T$. This analysis is
appropriate to the case where both grain and substrate arre insulating.
We then go on to include electrons - in this case phonons become
irrelevant, and only the combination of electrons and nuclei are important.

Before discussing the details, let us first state the method we shall use.
We have already seen how the dynamics of $\vec{S}$ is given in the
presence of a spin bath (Eqs.(\ref{4.23}) or  (\ref{4.38})). When we
couple in a bath of oscillators which allow transitions between
states $\mid \vec{S}_1; \chi_1 \{ \vec{\sigma}_k \} \rangle$ and
$\mid \vec{S}_2; \chi_2 \{ \vec{\sigma}_k \} \rangle$ of the
combined grain/spin bath system, in which there is no restriction
on the difference $2M$ between the polarisations of the spin bath states
$ \chi_1  \rangle$ or $ \chi_2  \rangle$, or on their biases
$\epsilon_1$ and $\epsilon_2$ (apart from those imposed by energy
conservation), the form of $P(t)$ must change. In this paper we will not
attempt to give a general expression covering all mechanisms simultaneously;
this is more than we need. Instead we will use 2 expressions, which apply
in the 2 limiting cases of interest. These are

(a) For short times, when the spin-mediate relaxation dominates; then
we ignore the oscillator bath, and use Eq.(\ref{4.10}).

(b) For longer times the oscillator bath-mediated relaxation will take
over. In this regime, each time $\vec{S}$ flips, some of the bias
energy is taken up by an oscillator mode, and the rest by some
number $r$ of flipped spins in the spin bath. There is no restriction
on either $r$ or the polarisation change $2M$, for this process
to work. The complete calculation of $P(t)$ here is very complicated,
but we notice that as soon as we reach times where the typical
oscillator-mediated transition rate, at a typical  bias, is
faster than the spin bath-mediated transition rate (for those grains
in an ensemble which have not yet relaxed), then a reasonable approximation
for $P(t)$ will be
\begin{equation}
P(t;T,\xi_H) =\int d\epsilon {W(\epsilon-\xi_H )
 \over Z(\beta ) } e^{-\beta (\epsilon-\xi_H) } \bigg\{ P^{(eq)}(T,\epsilon )
+\big[1- P^{(eq)}(T,\epsilon ) \big]  e^{-t/\tau(\epsilon ,T)} \bigg\}  \;,
\label{5.1}
\end{equation}
where $\tau^{-1} (\epsilon , T)$ is the oscillator-mediated relaxation rate,
and $P^{(eq)}(T,\epsilon )=e^{-\epsilon /T} /(2 \cosh (\epsilon /T) )$ is
the equilibrium population of the state $\vec{S}_1$ in a given bias
(in our previous discussion of the nuclear spin effects we assumed that
this bias was actually much less than $T$).
We note that this expression is {\it incoherent}, because it applies to
the majority of grains in a large bias, which cannot relax via  the
spin bath as discussed in the previous Section. The relaxation
rate in this incoherent case can be calculated as a second order
perturbation theory expression in the tunneling amplitude $\Delta_o$.
Thus the only effect of the spin bath which is left is the distribution
over the bias.

The formal proof of the above statement is easy to follow for the case
of pure othogonality blocking. In a typical bias $\epsilon \sim
\omega_o N^{1/2} \gg \Delta_o$ the second order perturbation theory
expression for the tunneling rate is given by
\begin{equation}
\tau^{-1}(\epsilon ) =2\pi \sum_{if} \rho_i^{eq} \mid U_{if} \mid ^2
\sum_{\alpha \gamma} \rho_\alpha ^{eq} \mid V^{(ph)}_{\alpha \gamma} \mid ^2
 \delta \big(\epsilon +E_{\alpha} -E_{\gamma} +\omega_o(M_i-M_f)/2 \big) \;,
\label{5.1a}
\end{equation}
where the sums are over the initial and final states of the spin bath
$(i,f)$ and oscillator bath $(\alpha ,\gamma )$ with the equilibrium
density matrices $\rho^{eq}$. Here $V^{(ph)}$ describes the interaction
with the phonons (we do not even need its explicit form to prove the point),
and the orthogonality rotation operator $U$ is given by
\begin{equation}
U=\prod_{k=1}^N e^{-i\beta_k \hat{\sigma}_k^y} \;.
\label{5.1b}
\end{equation}
The crucial point here is that in a large bias one may drop the
energy $\omega_o(M_i-M_f)/2$ transferred to the spin bath  from the
$\delta $-function argument in (\ref{5.1a}), because
only transitions for which $M_i-M_f \ll N^{1/2}$
contribute to the answer. After that
we have the sum over the complete set of states $f$, so that
 $U_{if}U^{\dag}_{fi} \equiv 1$ because the of the unitarity of $U$,
and the final answer is that for the oscillator bath alone.
Including topological decoherence changes the result in a minor way.
Indeed, with nonzero $\alpha_k$, we have
\begin{equation}
U=\prod_{k=1}^N e^{-i\beta_k \hat{\sigma}_k^y} \left(
e^{i\Phi+i\alpha_k \vec{n}_k \cdot \hat{\vec{\sigma}}_k }+
e^{-i\Phi-i\alpha_k \vec{n}_k \cdot \hat{\vec{\sigma}}_k }
\right) \;,
\label{5.1c}
\end{equation}
and the result for the relaxation rate in the oscillator bath
is renormalized by the factor
\begin{equation}
D=\langle U U^{\dag} \rangle =2+2\cos (2\Phi ) F_{\lambda}(1) \;,
\label{5.1d}
\end{equation}
Unless $\lambda =0$ and the topological phase  $\Phi $
is a multiple of $\pi$, this factor is of order $\sim 1$, and clearly has no
essential effect on the results.
Thus, apart from the renormalisation factor $D$, the relaxation
of a single grain in a nuclear bias field $\epsilon$ is that already
given in sections \ref{sec:2} and \ref{sec:3}, and all that remains for
a grain ensemble is to average over the bias field.

\subsection{Spin Bath Plus Phonons}

The case of insulating grains and an insulating substrate is the one where
there is most obvious competition  and interplay, between the nuclear
spin and phonon relaxation mechanisms. They are both slow.
We have already seen (Fig.6, and Eq.(\ref{4.43})) what the short-time
relaxation will look like. Now consider the implications of (\ref{5.1})
for the long-time relaxation. We worked out $\tau^{-1}(\epsilon )$
for phonon relaxation in the absence of nuclear spins; for a longitudinal
bias this was given by (\ref{3.13}), and will be roughly the same in the
presence of the spin bath (i.e., with the renormalisation factor (\ref{5.1d})
above). Thus we write $\tau^{-1}(\epsilon ,T)$ in the form
\begin{equation}
\tau^{-1}(\epsilon ,T) = \tau^{-1}_o
\left( { \epsilon \over E_o } \right) ^3 \coth (\epsilon /2k_BT )    \;,
\label{5.6}
\end{equation}
where $\tau^{-1}_o$ is a "typical" normalising relaxation rate, defined as
\begin{equation}
\tau^{-1}_o = \tau^{-1}(\epsilon = E_o, k_BT =E_o)  \sim
S^2 \Delta_o \left( {\Delta_o \over \Theta_D } \right)
\left( { E_o \over \Theta_D } \right) ^3 \;,
\label{5.7}
\end{equation}
and $E_o$ is the width of the Gaussian peak in $W(\epsilon )$, i.e.,
\begin{equation}
E_o =\omega_o N^{1/2} /2 \;.
\label{5.8}
\end{equation}
Notice that for a fixed bias $\epsilon =E_o$, the actual relaxation rate
$\tau^{-1}(E_o ,T)$ still depends on temperature as $\tau^{-1}(\epsilon ,T) =
\tau^{-1}_o \coth (E_o /2k_BT)$. At low $T$ this scales down to the constant
value $\tau^{-1}_o$, whereas at high $T$ it increases linearly, i.e.,
$\tau^{-1}(E_o ,T) = 2\tau^{-1}_o k_BT/E_o$.

It is useful to get some preliminary idea of typical time scales for $\tau_o$.
Let us consider 3 examples, viz., (i) a Tb Oxide grain, with $S\sim 1000$,
containing $N \sim 1000$ nuclei; (ii) a particle of Er$\:$As (such particles
are apparently insulating \cite{30}, containing  $N \sim 1000$ nuclei, with
$S \sim 1000$; and (iii) Ni$\:$O grains with $S \sim 1000$, but $N \sim 10$
only (coming from the $1\% $ of Ni$^{59}$ nuclear spins). Let us also
assume that $\Theta_D \sim 100\:K$ and $\Delta_o \sim 1\:MHz$ for each
example.  However $E_o$ varies dramatically  - for the Tb$\:$O grain,
$\omega_o \sim 5\:GHz$, and $E_o \sim 4\:K$; for the Er$\:$As grain,
$\omega_o \sim 1\:GHz$, and $E_o \sim 0.8\:K$; whilst for the Ni$\:$O
grain, $\omega_o \sim 28\:MHz$, and $E_o \sim 2.3\times 10^{-3}\:K$.
We thus get a wide range of time scales; for the Tb$\:$O grains one has
$\tau_o \sim 1.6 \times 10^{-2}\:s$, for the Er$\:$As grains, one has
$\tau_o \sim 5\: s$, whilst for the Ni$\:$O grains, one has
$\tau_o \sim 4.5\:$years!

We now consider the general behaviour of $P(t,T;\xi_H )$. We start by
noticing that the weighting function for the initial states, of bias
$\epsilon$, contributing to (\ref{5.1}), will be
\begin{eqnarray}
w(\epsilon , T; \xi_H ) &=& W(\epsilon - \xi_H ) {e^{-\beta (\epsilon -\xi_H )}
\over Z(\beta ) }  \nonumber \\
& \sim & { 1 \over \sqrt{2 \pi} E_o } \exp \bigg\{
-{1 \over 2} \bigg[ {\epsilon -\xi_H \over E_o} + {E_o \over k_BT } \bigg]
\bigg\} \;.
\label{5.9}
\end{eqnarray}

Ignoring $\xi_H$ for the moment, we see that in zero external field this
distribution peaks at {\it negative} bias $\epsilon \sim -E_o^2/k_BT$.
This is natural; as we lower $T$, more and more of the nuclei align
with $\vec{S}$, thereby lowering the energy of the combined system in the
initial state. This process continues until $\sim N$ nuclei are aligned with
$\vec{S}$, i.e., when $\epsilon \sim -N\omega_o/2$, at a temperature
$k_BT \sim \omega_o /2$. We have then reached the bottom edge of the
distribution $W(\epsilon )$, way out in the wings. If $\xi_H \ne 0$, the
distribution peaks at $\epsilon \sim (\xi_H -E_o^2/k_BT)$.

Thus we write
\begin{eqnarray}
P(t,T;\xi_H) &=&\int d\epsilon w(\epsilon , T; \xi_H )
 \bigg\{ P^{(eq)}(T,\epsilon )
+\big[ 1- P^{(eq)}(T,\epsilon ) \big]
e^{-t/\tau(\epsilon ,T)} \bigg\} \nonumber \\
&=& \int { dx \over \sqrt{2 \pi} }
e^{-{1 \over 2}(x-\overline{\xi}_H+1/\overline{T})^2}
 { e^{-x/\overline{T}}+e^{x/\overline{T}}
e^{-(t/\tau_o)\:x^3 \coth (x/2\overline{T}) } \over 2\cosh (x/\overline{T}) }
\;,
\label{5.10}
\end{eqnarray}
where we have normalised all energies by $E_o$:
\begin{eqnarray}
x &=& \epsilon /E_o
 \nonumber \\
\overline{\xi}_H &=&\xi_H /E_o
\label{5.11} \\
\overline{T} &=& k_BT /E_o  \;. \nonumber
\end{eqnarray}

Let us first look at the high-$T$ limit of (\ref{5.10}), i.e., for
$k_BT \gg E_o$ when $\overline{T} \gg 1$. Then (\ref{5.10}) is easily
evaluated; one finds
\begin{eqnarray}
P(t,T;\xi_H)-1/2 &\sim &
{ 1 \over 2}{1 \over  (1+4\overline{T}t / \tau_o )^{1/2} }
\exp \bigg\{ -\overline{\xi}_H^2 {2\overline{T}t \over
 \tau_o+4\overline{T}t} \bigg\}
\label{5.12} \\
&\longrightarrow &
\; { 1 \over 2}{1 \over  (1+4\overline{T}t / \tau_o )^{1/2} }\;;\;\;\;
\;\;\;\;\; (k_BT \gg E_o )\;.
\label{5.13}
\end{eqnarray}
(We recall that this result is correct only when $T \ll \Omega_o$ -
otherwise higher levels of the grain will come into play.)

This surprising power-law behaviour is actually easily understood.
The initial fast relaxation
(at $t \ll \tau_o$) comes from  grains with large
bias; the slower  relaxation
($t \gg \tau_o$) comes from the grains with smaller bias.
The typical relaxation occurs at a rate $\overline{T}\tau^{-1}_o$, i.e.,
it is faster for higher temperatures, in accordance with the remarks just
after Eq.(\ref{5.8}). We may think of the power law as a "grain-ensemble"
sum of a lot of different exponential decays, or as a funny kind of
"stretched exponential". Note however that the decay is {\it not}
logarithmic in time.

Now consider the low-$T$ limit, with $\overline{T} \ll 1$; this is a little
more complicated. From (\ref{5.10}) we now have
\begin{eqnarray}
P(t,T;\xi_H) &\approx&1+ \int { dx \over \sqrt{2 \pi} }
e^{-{1 \over 2}(x-\overline{\xi}_H+1/\overline{T})^2}
 { e^{x/\overline{T}} \over 2\cosh (x/\overline{T}) }
\left( e^{- x^3 \coth (x/2\overline{T} )
   t/\tau_o } -1 \right)  \nonumber \\
&=&1+e^{ \overline{\xi}_H/\overline{T} - 1/2\overline{T}^2  }
\int { dx \over \sqrt{2 \pi} }
{ e^{-(x-\overline{\xi}_H)^2/2 }  \over 2\cosh (x/\overline{T}) }
\left( e^{- x^3\coth (x/2\overline{T} )  t/\tau_o } -1 \right) \;,
\label{5.14}
\end{eqnarray}
Let us consider first the behaviour of (\ref{5.14}) when
$ \overline{\xi}_H =0$. Using the fact that $\overline{T} \ll 1$ one may
further simplify this expression  to
\begin{equation}
P(t,T;\xi_H=0) \approx 1+ { \overline{T} \over \sqrt{2 \pi} }
e^{ - 1/2\overline{T}^2  }
\bigg[ \int_0^\infty { dx   \over \cosh (x) }
 e^{- x^3 \coth (x/2 ) \: t\overline{T}^3/\tau_o } -{\pi \over 2}\bigg] \;,
\label{5.15}
\end{equation}
This clearly defines a temperature-dependent relaxation rate
\begin{equation}
\tau^{-1}_{eff}(T) = \overline{T}^3\tau_o^{-1} \sim  \;
S^2 \Delta_o \left( {\Delta_o \over \Theta_D } \right)
\left( { k_BT \over \Theta_D } \right) ^3 \;,
\label{5.16}
\end{equation}
which goes to zero at low $T$. We note however that the validity of
(\ref{5.15}) requires that the calculation is not affected by the edges
of $W(\epsilon )$, which means that at a temperature $k_BT \sim \omega_o/2$,
Eq.(\ref{5.16}) crosses over to a constant value, i.e.,
\begin{equation}
\tau^{-1}_{eff}(T \to 0) \rightarrow N^{3/2}\tau_o^{-1} \sim
S^2 \Delta_o \left( {\Delta_o \over \Theta_D } \right)
\left( { \omega_o \over 2\Theta_D } \right) ^3  \;.
\label{5.17}
\end{equation}
but the magnitude of the relaxing component at this temperature is already
extremely small; in fact  $1-P(t \to \infty ) \sim N^{-1/2}e^{-N}$.
It is again useful to consider what is the
the maximum relaxation time at low $T$. For the examples previously
mentioned, we have (i) for the Tb$\:$O grain, $\tau_{eff} (T \to 0)
\sim 10$ minutes; (ii) for the Er$\:$As grain, $\tau_{eff} (T \to 0)
\sim 2$ days; and (iii) for the Ni$\:$O grain,
$\tau_{eff} (T \to 0) \sim 4\times 10^7$ years.

Defining the exponentially small
total amplitude of the relaxing component as
\begin{equation}
A \approx { \overline{T}\sqrt{\pi } \over 2\sqrt{2} }
e^{ - 1/2\overline{T}^2  } \;,
\label{5.18}
\end{equation}
we find the long time asymptotics of $P(t)$ as
\begin{equation}
P(t,T;\xi_H=0) \longrightarrow  1-A + A
\left(  { 2\tau_{eff}(T) \over \pi t} \right)^{1/2} \;.
\label{5.19}
\end{equation}

The reason for the $T^{3}$ decrease in the relaxation rate at low $T$
is simply that as we lower the temperature, the typical bias energy
decreases linearly with $\overline{T}$, because only a small fraction
of grains in the Gaussian tail which have very small bias $\epsilon
\sim k_BT$ contribute to the relaxation. The majority of the grains
are actually {\it "trapped"}
by the hyperfine interaction in a negative bias energy
$\epsilon \sim E_o^2/k_BT \gg k_BT$. Thus at this low temperature
even phonons will not help to liberate the giant spin.

Consider now Eq.(\ref{5.14}) with nonzero external bias.
For $\overline{\xi}_H \ll 1/\overline{T}$ the previous answer (\ref{5.19})
hardly changes except that the amplitude of the relaxing component is given
now by
\begin{equation}
A \approx { \overline{T}\sqrt{\pi } \over 2\sqrt{2} }
e^{ - 1/2(\overline{\xi}_H -1/ \overline{T} )^2  } \;,\;\;\;\;
(\overline{\xi}_H \ll 1/\overline{T})\;.
\label{5.20}
\end{equation}
We observe that already for rather small bias $\overline{\xi}_H
 \sim \overline{T} \ll 1$ we have an exponential dependence on
$\xi_H$. The negative bias will simply further suppress an already
exponentially small fraction of relaxing grains and we shall
not go into more details here. The case of positive $\xi_H$ is
much more intriguing and surprising. After an exponential
increase in the amplitude up to $A \sim 1$ for
$\overline{\xi}_H \sim 1/\overline{T}$, the answer changes drastically.
For $\overline{\xi}_H > 1/\overline{T} $ one finds the magnetisation function
to be
\begin{equation}
P(t,T;\xi_H) \approx   \int { dx \over \sqrt{2 \pi} }
e^{-{1 \over 2}(x-\overline{\xi}_H+1/\overline{T})^2}
 e^{-\mid x \mid ^3\: t/\tau_o } \;,\;\;\;(\overline{\xi}_H>1/\overline{T}) \;,
\label{5.21}
\end{equation}
which has different behaviour depending on whether we look at
short times, where
\begin{eqnarray}
P(t,T;\xi_H) &\approx& e^{-t/\tau_{eff}(\xi_H,T) } \;, \;\;\;\;
(t/\tau_{eff} < ( \overline{\xi}_H-1/\overline{T} )^2 )\;,
\label{5.22}  \\
\tau_{eff}^{-1}(\xi_H,T)& =&
( \overline{\xi}_H-1/\overline{T} )^3 \tau_o^{-1} \;;
\label{5.23}
\end{eqnarray}
or long times, where
\begin{equation}
P(t,T;\xi_H) \sim {\sqrt{2} \Gamma (1/3) \over 3 \sqrt{\pi } }
e^{-( \overline{\xi}_H-1/\overline{T} )^2/2 }
\left( {\tau_o \over t } \right)^{1/3} \;;\;\;\;\;
(t/\tau_{eff} > ( \overline{\xi}_H-1/\overline{T} )^2 )\;.
\label{5.24}
\end{equation}
Note that Eqs.(\ref{5.21})-(\ref{5.24}) are derived in the approximation
that $ \overline{\xi}_H-1/\overline{T} \gg 1 $. As before, the bias is
restricted to be much less than $\Omega_o$.
Finally, at $t >\tau_o$, this behaviour changes yet again to
$P(t) \approx \exp \{-( \overline{\xi}_H-1/\overline{T} )^2/2 \}\;
(\tau_o/4t\overline{T} )^{1/2}$.
We note
the anomalous temperature and bias dependence of the effective relaxation rate
(\ref{5.23}) near the crossover.

{}From this analysis we conclude that unless the external bias is
larger than $E_o$ (or  $E_o^2/k_BT$ at low temperature) the ensemble
averaged results for the magnetisation relaxation {are not described}
by the naive theory of a two level system coupled to a bath of oscillators,
and spin bath interactions essentially modify the answer. It is worth
noting three effects which have to be kept in mind when considering the
evolution of the experimental data in external bias:

(i) Since the dependence on $\xi_H$ starts when
$\xi_H > E_o$ or  $\xi_H > E_o^2/k_BT$, one {\it cannot} derive
the value of the tunneling amplitude from this dependence, as it
would be in the case of isolated two-level system;

(ii) At low temperature the giant spin is "trapped" in its initial state
by the large and negative bias produced by the spin bath.

(iii) Only large and positive bias can liberate the giant spin and to
allow complete magnetisation relaxation. The temperature and bias dependence
of the relaxation rate is anomalous near the crossover region
 $ \overline{\xi}_H \sim 1/\overline{T}$. Of course, when
  $ \overline{\xi}_H \gg 1/\overline{T}$
  we  recover back the pure case
of oscillator bath relaxation.

At this point we would like to comment on our starting
assumption that the initial
state of the spin bath is equilibrated with the giant spin direction
$\vec{S} =\vec{S}_1$. Experimentally this could be arranged by applying a
very strong negative bias $\xi_H$ during a time period much longer
than the longitudinal  NMR relaxation time $T_1$, and switching it off
at $t=0$. In some systems however this procedure may not work because
of an  astronomically long $T_1$ (in fact,
low temperature longitudinal NMR is still something of an unsolved mystery;
spin-lattice relaxation times at $mK$ temperatures ought to be many years
in insulating crystals, unless some gapless magnetic excitations are
involved, e.g., on the sample surface  or  crystal defects).
It may be useful then to introduce two {\it different} temperatures -
one for the oscillator bath $T_{ph}$ and the other for the spin bath $T_S$
(which may even be negative, i.e.,  $T_S <0$!).
Our basic equation (\ref{5.1})
is still valid in this more general case, but now the distribution
over the initial bias is defined by the spin bath temperature, whereas the
giant spin evolution toward equilibrium is governed
by the crystal temperature, i.e.,
\begin{equation}
P(t,\xi_H) =\int d\epsilon {W(\epsilon-\xi_H )
 \over Z(T_S ) } e^{-(\epsilon-\xi_H)/T_S } \bigg\{ P^{(eq)}(T_{ph},\epsilon )
+\big[1- P^{(eq)}(T_{ph},\epsilon ) \big]
e^{-t/\tau(\epsilon ,T_{ph})} \bigg\}  \;.
\label{5.25}
\end{equation}
One may proceed with the analysis of this expression as before.
At this point we feel that considering more cases in this paper will
not add much to the physical picture. Depending on the particular
experimental system and sample preparation, the necessary formulae  can
be easily derived from (\ref{5.25}).

Summarising what we have found for insulating systems, we see that at
short times, the relaxation proceeds entirely via the nuclear spin bath,
and is logarithmic in time (Eq.(\ref{4.43})). At longer times phonons
take over, and we get power law decay in time; at low $T$ this
goes as $(\tau_{eff}(T)/t)^{1/2}$, with $\tau_{eff}^{-1} \sim T^3$
(Eqs.(\ref{5.16}) and (\ref{5.19})), but only a small fraction
of grains relax unless the bias $\xi_H \sim E_o/k_BT $ or larger. This bias
is necessary to counteract the nuclear bias field. At higher temperatures
we still get power-law relaxation (Eq.(\ref{5.13})). The crossover
between nuclear spin-mediated and phonon-mediated relaxation can be
understood by matching Eq.(\ref{4.43}) with the relevant phonon expression.

\subsection{Spin Bath Plus Electrons}

We now turn to the case where either the grain or the substrate is
{\it conducting}. We use the giant Kondo model of section \ref{sec:2}
to describe interaction of $\vec{S}$ with electrons, with a dimensionless
coupling $\alpha$. If both grain and substrate are conducting, then
$\alpha =\alpha_b \sim g^2S^{4/3}$ (Eq.(\ref{2.7})); if only the substrate
conducts, then $\alpha =\alpha_s \sim g^2 S^{2/3}$ (Eq.(\ref{2.13})); and
if only the grain conducts, $\alpha \sim \alpha_s
e^{-\Delta \!\epsilon_{\mu} /T}$
(Eq.(\ref{2.21})).

The formal analysis is almost identical to that just used for phonons -
we start again from (\ref{5.1}), now using the general result (\ref{2.8})
for the electronic relaxation rate. The high temperature limit
is most transparent because $\tau_e(T)$ in Eq.(\ref{2.9})
is independent of the bias. Thus we find a pure exponential relaxation
\begin{equation}
P(t)=1/2 \big( 1+e^{-t/\tau_e(T)} \big) \;; \;\;\;\;(\alpha T, T \gg E_o)\ ;.
\label{5.26}
\end{equation}
at high temperature.

For very small grains or conducting grains on an
insulating substrate at $T \ll \Delta \! \epsilon_{\mu}$, we have
a peculiar relaxation regime when the internal bias is much less than
temperature, but the electronic damping rate $\Gamma_e =2\pi \alpha T$
 is already small, that is for $\alpha \ll 1$ there is the
temperature range where $\Gamma \ll E_o \ll T$. Now the electronic
relaxation rate is inversely proportional to the bias energy
\begin{equation}
\tau_e^{-1}(T,\epsilon) = 8\pi \alpha T {\Delta_o^2 \over \epsilon^2 }\;,\;\;\;
(2\pi T \ll \epsilon  \ll T)\;.
\label{5.27}
\end{equation}
Substituting this expression to (\ref{5.1}) we find the time correlation
function as
\begin{equation}
P(t,T;\xi_H) \approx {1 \over 2} +{1 \over 2} \int { dx \over \sqrt{2 \pi} }
e^{-{1 \over 2}(x-\overline{\xi}_H)^2}
 e^{- t/(x^2\tau_e(T,E_o) }   \;,
\label{5.28}
\end{equation}
which after easy evaluation yields
\begin{eqnarray}
P(t)&=& 1/2 \big( 1+e^{-\sqrt{2t/\tau_e(T,E_o)}} \big) \;; \;\;\;\;
(\xi_H =0)
\label{5.29}  \\
& \approx & 1/2 \big( 1+e^{-t/\tau_e(T,\xi_H)} \big) \;; \;\;\;\;\;\;\;\;
(\xi_H \gg E_o)\;.
\label{5.30}
\end{eqnarray}
The temperature dependence of the effective relaxation rate is
$ \sim T\alpha(T)$, and again in the small external bias we have an unusual
decay law - it is neither simple exponential nor power-law.

Let us deal now with the low-temperature behaviour, $k_BT/E_o \ll 1$.
Using (\ref{2.10}) we then get
\begin{eqnarray}
P(t,T;\xi_H) &\approx & 1+ \int { dx \over \sqrt{2 \pi} }
e^{-{1 \over 2}(x-\overline{\xi}_H+1/\overline{T})^2}
 { e^{x/\overline{T}} \over 2\cosh (x/\overline{T}) }
\left( e^{-f(x)\; t/\tau_{eff} } -1 \right)  \;,
\label{5.31} \\
f(x)&=&\cosh (x)\: \mid \Gamma [\alpha +ix/\pi ]\mid^2 \; \nonumber
\end{eqnarray}
(compare (\ref{5.14})).
If the bias $\overline{\xi}_H < 1/\overline{T}$,
this defines a temperature dependent relaxation rate
\begin{equation}
\tau_{eff}^{-1}(T) = { 2 \over \Gamma [2\alpha] }
  \Delta_o \left( {\Delta_o \over \Omega_o } \right)
\left( {2\pi T \over \Omega_o } \right)^{2\alpha-1} \;.
\label{5.32}
\end{equation}
The answers
crucially depend on the parameter $\alpha$. We consider three limiting cases:

(i) For $\alpha \ll 1$ we approximate the function $f(x)$
by $f(x) \approx x \coth (x) (\alpha^2 + (x/\pi)^2 )^{-1}$ and obtain
\begin{eqnarray}
P(t,T;\xi_H) &\approx & 1-A{\pi t \over \alpha \tau_{eff} }\;;
\;\;\;\;\;(t \ll \alpha^2 \tau_{eff})  \;,
\label{5.33} \\
P(t,T;\xi_H) &\approx & 1-2A\left( {\pi t \over \tau_{eff} } \right)^{1/2} \;;
\;\;\;\;\;( \alpha^2 \tau_{eff} < t < \tau_{eff}) \;,
\label{5.34} \\
P(t,T;\xi_H) &\approx & 1-A +A \left( { t \over \tau_{eff} } \right)^{1/4}
e^{-2\pi \sqrt{t/\tau_{eff} } } \;;
\;\;\;\;\;( \tau_{eff} \gg t) \;.
\label{5.35}
\end{eqnarray}

(ii) For $\alpha =1/2$ we have $f(x)=\pi $, and, as in the high temperature
limit, the relaxation is given by the  simple exponential law
\begin{equation}
P(t)=1-A \big( 1-e^{-\pi t/\tau_{eff} } \big) \;;
\label{5.36}
\end{equation}

(iii) Finally we consider the case of large $\alpha$, which has a power-law
asymptotic at long times
\begin{equation}
P(t,T;\xi_H) \approx  1-A +A 2 \Gamma \big[ { 2\alpha \over 2\alpha -1} \big]
 \left( { \tau_{eff} \over \pi t } \right)^{1/(2\alpha -1)}
\label{5.37}
\end{equation}
In all these expressions  $A(T,\xi_H)$ is given by (\ref{5.20}).

The interpretation of these results also depends on the value of $\alpha$.
If both grain and substrate are conducting, then the relaxation time in
(\ref{5.32}) will be greater than the age of the universe unless $S \le 100$
(depending on the value of $g^2$). This shows the astonishing
power of the electron bath to "freeze" the dynamics of $\vec{S}$, unless
it is quite microscopic in size. We shall not analyze
here the case of very small $S$, since once $S \sim O(10-100)$, the detailed
structure of $W (\epsilon )$ will no longer be Gaussian - there will be lots
of fine structure, depending on the particular systen involved. Such studies
are best done on a case-by-case basis.

If only the substrate is conducting, then $\alpha =\alpha_s$ will be small
until $S \sim 10^3$. For larger $S$ then the giant spin is again frozen.
For $S \le 10^3$ the results  (\ref{5.33})-(\ref{5.36}) can be applied
if $\overline{\xi}_H \ll 1/\overline{T}$. Again, as
$\overline{\xi}_H $ approaches $ 1/\overline{T}$, there will be
an amusing crossover , with a relaxation rate given for finite
$\overline{\xi}_H > 1/\overline{T}$ by
\begin{equation}
\tau_{eff}^{-1}(T, \xi_H) =\tau_e^{-1}(T,\epsilon =  \xi_H-E_o^2/T) \;,
\label{5.38}
\end{equation}

Finally, if only the grain is conducting,
$\alpha$ can be exponantially small, if the spacing $\Delta \!\epsilon_{\mu}$
between the internal electron levels is large enough - this typically
requires that $\Delta \!\epsilon_{\mu}\ge \Omega_o$, so that $S \le 10^5$
(cf. Eq.(\ref{2.23})). In this case, unlike the cases where the substrate
is conducting, the short-time relaxation will be again dominated
by nuclear spin-mediated transitions, and the weak electron-mediated
transitions only enter at later times in possible competition with the
phonon-mediated transitions; the typical electron-mediated rate becomes
\begin{equation}
\tau_{eff}^{-1}(T)=4\pi \alpha_s {\Delta_o^2 \over E_o^2}
e^{-\Delta \!\epsilon_{\mu}/T} \;,
\label{5.39}
\end{equation}

Note that for low enough temperature, $\tau_{eff}$ in (\ref{5.39})
becomes so long that electrons in the grain become irrelevant, and we
return to the calculations for the nuclear spin/phonon-mediated
problem, i.e., the grain behaves as an insulator.

Thus to summarize, we see that depending on whether the grain and/or
substrate are conducting, we get behaviour ranging from grains frozen
for all eternity (if $\alpha \gg 1$), to grains behaving essentially
as insulators, with short-time relaxation controlled by the nuclei
(when $\alpha \to 0$). In the former case we see an extraordinary
alliance between the electrons  (with their strong dissipative suppression
of tunneling) and nuclear spins (which trap $\vec{S}$ in a negative bias
$\epsilon \sim -E_o^2 /T$) to block any motion of $\vec{S}$ even at low
$T$, where traditionally one expects tunneling. In fact, as $T \to 0$,
our calculations show that $\vec{S}$ is blocked for eternity
when $S$ takes microscopic values (e.g., $S \sim O(10)$), because the
negative bias becomes very large. This is a quite astonishing demonstration
of the control that nuclear spins exert over the dynamics of $\vec{S}$ -
the spin is only unblocked when $S \sim O(1)$, when $\alpha \ll 1$ again
(and the whole WKB framework breaks down!).

\section{Summary \& Conclusions: Practical Implications}
\label{sec:6}

Let us first summarize our main results. At temperatures
$T$ such that $k_BT \ll \Omega_o$, we may employ a 2-level model
with bias to describe the grain. Coupling to electrons and/or
phonons in this model can be treated by the usual oscillator
bath methods. However the crucial coupling turns out to be to the
nuclear spin bath, to which oscillator bath models do not apply.
The main effect of this coupling is to spread each of the 2 levels
for $\vec{S}$ into a Gaussian "band" of half-width $E_o \sim N^{1/2}
\omega_o$, and total width $N\omega_o$, where $\omega_o$ is the
hyperfine coupling and $N$ the number of nuclei in the grain.
If $N$ and $\omega_o$ are small, $E_o$ may only be a few $mK$
(as in Ni or Fe grains). If $N$ and $\omega_o$ are large (as in mesoscopic
rare earth grains), $E_o$ may be hundreds of Kelvin or more
(so that $E_o \gg \Omega_o$). At low temperatures the nuclei
begin to line up with the grain vector $\vec{S}$, and then $\vec{S}$
finds itself in a {\it negative} internal nuclear bias field, with a mean
bias $\epsilon$ of roughly $\epsilon \sim -E_o^2/T$. Then at short
times the only way that the grain can relax is by taking energy from the
nuclear system. The inter-nuclear dipolar interactions allow this to happen,
by causing the bias $\epsilon$ to become time-dependent - the result
is a slow relaxation, which for an ensemble of grains gives a fraction
of relaxed grains going roughly logarithmically in time (section \ref{sec:4}):
\begin{equation}
1-P(t) \sim  \sqrt{ {1 \over2\pi  N} }
 { \ln \left( 2\tilde{\Delta}_o^2t/\pi^{1/2} \Gamma_\mu  \right) \over
 \ln \bigg[ {1 \over e\sqrt{\lambda } } \ln
\left( 2\tilde{\Delta}_o^2t/ \pi^{1/2} \Gamma_\mu \right)  \bigg] }   \;.
\label{6.1}
\end{equation}
This logarithmic relaxation, shown in Fig.7, has nothing to do
with a distribution of sizes of the grains (we assume all grains have the
{\it same} parameters).

At longer times the grains can relax very slowly via either electron-
or phonon-mediated transitions, again in the time-varying nuclear field.
If the grain is insulating (no electrons), we get power-law
relaxation; a fraction $A$ of grains can relax, where
\begin{equation}
A \approx \sqrt{\pi/2 } { T \over 2E_o }
\exp \bigg\{ - 1/2\left(
{\xi_H  \over E_o} -{E_o  \over T}  \right)^2  \bigg\} \;.
\label{6.2}
\end{equation}
in a bias field $\xi_H$, and these grains relax as $A (\tau_{eff} /t)^{1/2}$
where
\begin{equation}
\tau^{-1}_{eff}(T)  \sim
S^2 \Delta_o \left( {\Delta_o \over \Theta_D } \right)
\left( { k_BT \over \Theta_D } \right) ^3 \;,
\label{6.3}
\end{equation}
if $\xi_H \ll E_o^2/T$, and $P(t) \sim 1/2(1+e^{-t/\tau_{eff}})$ with
\begin{equation}
\tau_{eff}^{-1}(\xi_H,T) =
\left(  {\xi_H  \over E_o} -{E_o  \over T}   \right)^3 S^2 \Delta_o
\left( {\Delta_o \over \Theta_D } \right)
\left( { E_o \over \Theta_D } \right) ^3 \;,
\label{6.4}
\end{equation}
once $\xi_H > E_o^2/T$. Thus very few grains can relax at until the
external bias $\xi_H$ compensates the internal bias $\epsilon$ and
"untraps" $\vec{S}$.

If there are electrons around, the results depend on whether the
substrate is conducting or not. If it is, then all but microscopic
spins ($S \ll 100$) will be trapped in states of frozen magnetisation
for astronomical times, once the temperature goes below $\Omega_o$.
This remarkable result is a combination of "degeneracy blocking"
caused by the nuclear bias field, and the very strong electronic
dissipation. However if only the grain is conducting, and the
substrate is insulating, then for $S \ll 10^5$, the electronic
relaxation becomes negligible, and we go back to the fully insulating case;
the full results for all values of electronic coupling appear in
section \ref{sec:5}.$\:$B.

It is clear that these results have implications both for future
magnetic device design, and for the thousands of experiments that have
been done on magnetic grain relaxation over the years. As far as
devices are concerned, perhaps the most interesting result is the
freezing of $\vec{S}$ when both grain and substrate are conducting
unless $S \ll 100$. This may have far-reaching implications for future
computers and information storage, since it implies that if we are prepared
to go to low temperatures, one should be able to use even magnetic
molecules as permanent memory storage elements, perhaps of the "M.R.A.M."
type (cf. ref.\cite{31}). Even more tempting is the possibility
that we may be able to manipulate $\vec{S}$ in such molecules
indirectly, by controlling the nuclear spin polarisation, since it is
the nuclear bias field that controls the dynamics of $\vec{S}$.

As far as experiments on grain relaxation are concerned, we
believe that it is quite urgent that experimental tests of our results
be done at low $T$. Unfortunately most of the many relaxation
measurements that have already been done are at higher temperatures,
where relaxation is dominated by thermal activation. What we have found
here is that the low-$T$ behaviour does not look like the conventional
tunneling picture at all - there is no crossover to a temperature -
independent "tunneling relaxation" at temperatures $k_BT \ll \Omega_o$.
The low-$T$ behaviour of magnetic particles is not governed by the
external bias at all, unless it is very large; instead it is governed
by the {\it random distribution}
of internal nuclear bias fields. This renders previous theories of the
tunneling of magnetic particles somewhat meaningless, regardless of the
external field. Thus we believe that it would be useful to look again
at those experiments that {\it have} been done at low $T$ on ensembles
of relaxing grains \cite{6,7,8,22,30}, particularly in an attempt
to understand the "plateau" that often appears at low $T$ in the
relaxation rate (the "magnetic viscosity plateau").

However an important cautionary note is necessary. In virtually all of
these experiments, {\it inter-grain interactions} (mediated by the grain
dipolar fields) are very important. We believe this to be the reason
for the common occurence of "avalanche" magnetisation reversal in grain
ensembles (see, e.g., ref.\cite{22}); these avalanches are {\it prima facie}
evidence for the importance of inter-grain interactions (such avalanches also
occur in multi-domain magnets, because of dipolar interactions between the
domains and walls \cite{7,32}). Thus to interpret most experiments
(including the remarkable recent results on Mn$_{12}$O$_{12}$ molecules
\cite{22}), we must incorporate these interactions - this we do elsewhere
\cite{23}.

However some experiments may not suffer this problem. The recent results
of Coppinger {\it et al}. \cite{30}, on ErAs grains, with $S \sim 10^3$,
may be a case in point. It is not yet clear what is the effective
coupling to the electron bath for this system, but we believe that the theory
in the present paper should be applicable to this case.

\section{Acknowledgement}
This work was supported by NSERC in Canada,
by the International Science Foundation (MAA300),
and by the Russian Foundation for
Basic Research (95-02-06191a).  We would also like to thank
I. Affleck, M. Bloom, B.G. Turrell, and W. Hardy for discussions
of the theory, and B. Barbara, F. Coppinger, D. Maude, C. Paulsen, and
J. Portal for discussions concerning experiments. More general
discussions concerning the spin bath, with
S. Coleman,  A.J. Leggett, R. Penrose, and W. Unruh, were also very
useful.

\appendix
\section{ Kondo-type renormalizations in the multi-orbital scattering}

In truncating the initial Hamiltonian (\ref{2.2}) to the 2-level system
for the giant spin one generates both the diagonal, i.e., proportional to
$\hat{\tau}_z  \hat{\sigma}_z $ term, and nondiagonal terms
$\sim \hat{\tau}_\pm  \hat{\sigma}_\mp $. The latter are usually not
present in the spin-boson Hamiltonian because they are subdominant to the
diagonal ones.
In general we may write
\begin{equation}
H^{eff}= \Delta_o \hat{\tau}_x +
\hat{\tau}_z  \hat{\sigma}_z^{\alpha \alpha} \sum_{kk^\prime l} J^z_l
c^{\dag}_{kl,\alpha} c_{k^\prime l,\alpha} +
\hat{\tau}_\pm  \hat{\sigma}_\mp^{\alpha, -\alpha} \sum_{kk^\prime l l^\prime }
 J^{\perp}_{ll^\prime }  c^{\dag}_{kl,\alpha} c_{k^\prime l^\prime, -\alpha}\;.
\label{a.1}
\end{equation}
Here $l$ and
$l^\prime$ are the orbital channels for electron scattering off the grain
($l_{max} \sim k_FR_o \gg 1$), and we write the Hamiltonian to be
diagonal in $l$ for $ \hat{\tau}_z  \hat{\sigma}_z $ coupling.
We observe that initially (before scaling) the transverse coupling
$J^{\perp}$ is exponentially small as compared with $J^z$. In fact
$J^{\perp}/J^z \sim \Delta_o/\Omega_o \sim \langle \Uparrow \mid \Downarrow
\rangle $ because transverse matrix elements flip the giant spin as well and
are therefore proportional to the overlap between the
$\mid  \Uparrow \rangle $ and $\mid  \Downarrow \rangle $ states.

Formally (\ref{a.1}) is a {\it strongly anisotropic} Kondo model with
multi-channel scattering in a {\it magnetic field} $\Delta_o$ applied along the
weak-scattering  direction. Since the $J^{\perp}$ are small we may ignore the
renormalization of the $J^z$ and study the renormalization group equation for
the $J^{\perp}$ first. Introducing dimensionless couplings
$g^{\perp}_{ll^\prime} = J^{\perp}_{ll^\prime} N(0)$ and the same for
$g^{z}_{l}$, we find
\begin{equation}
\dot{g}^{\perp}_{ll^\prime} = g^{\perp}_{ll^\prime} \bigg[
{\delta_l+\delta_{l^\prime} \over \pi} - \alpha_s \bigg]
\;,
\label{a.2}
\end{equation}
\begin{equation}
\alpha_s = {1 \over 2 \pi^2} \sum_l \delta_l^2 \;,
\label{a.3}
\end{equation}
where  $\delta_l$ is the scattering phase shift due to $g^z_l$.
The crucial point is that for a large grain, when many scattering channels are
equally coupled to the giant spin, the Anderson orthogonality \cite{6} term
$\alpha_s$ will always dominate in the scaling of $g^{\perp}$. Thus, quite
generally, we never face the strong coupling regime when the $g^{\perp}$ are of
order $g^z$ and large.

Naively one would ignore these transverse couplings completely, but we have to
remember that the giant spin dynamics is governed by the {\it transverse }
magnetic field $\Delta_o$ which is also subject to renormalization due to
Anderson orthogonality. The relevant renormalization group variable is fugacity
$\Delta_o /T$, and
\begin{equation}
\dot{\Delta}_s  = \Delta_s (1-\alpha_s ) \;,
\label{a.4}
\end{equation}
(the solution of this equation is explicitly used in our equations
(\ref{2.8})-(\ref{2.10})). Thus the question of whether we may neglect
conduction electron spin flips or not depends on the parameter
\begin{equation}
\theta = {\rm Max} \{ {\delta_l+\delta_{l^\prime} \over \pi} \} \;.
\label{a.5}
\end{equation}

If $\theta $ is less than unity we are back to (\ref{2.8})-(\ref{2.10}), and
electron spin flips are irrelevant. If we find $\theta >1$ then the correct
expression for the giant spin relaxation time will be
\begin{equation}
\tau^{-1} \approx 2\pi \mid J^{\perp}_{ll^\prime} \mid^2
N(0) \left( { (2 \pi T,\:\epsilon )_{max}  \over \Omega_o }
 \right)^{2\alpha_s-\theta } \;,\;\;\;
{ \delta_l+\delta_{l^\prime} \over \pi}=\theta
\label{a.6}
\end{equation}
Note that for $\theta >1$ we definitely have $\alpha_s \sim S^{2/3} \gg 1$,
so the $\vec{S}$ dynamics are definitely incoherent, and the above
refinement is hardly different from  (\ref{2.8})-(\ref{2.10}) (it is
 equivalent to going from $2\alpha_s-1$ to  $2\alpha_s-\theta$).

\vspace{2cm}
{\bf FIGURE CAPTIONS}

\figure{{\bf Figure 1} The Atomic Weight (in Daltons) of a typical computer
memory element, as a function of time. Note that the position
of the "Quantum Threshold" depends on temperature. As we see in this
article, it also depends strongly on what the element is made of.
 }

\figure{{\bf Figure 2} Schematic diagram of the energy scales in the giant
spin, biased by a longitudital field $H_z$. The bias $\epsilon =
\gamma_e S H_z$, and $\Omega_o$ is the gap to the next set of excited states.
In reality, for a giant spin, the top of the barrier will be much higher,
at energy $\sim (S\Omega_o )$ above the lowest states.
}

\figure{{\bf Figure 3} The distribution function $W(\epsilon )$ for spin
bath levels, around each giant spin level. Typically the parameter $\mu >1$
except for very small numbers of nuclear spins (see text).
}

\figure{{\bf Figure 4} Some possible transitions when $\vec{S}$ flips,
for a grain in zero external field. The initial state of the combined
giant spin/nuclear spin system has energy $\epsilon_1$; in the
diagram this state $\mid 1 \rangle$ has either initial nuclear polarisation
$\Delta \! N =M$ or  $\Delta \! N =M+2$. Transitions to states
$\mid 2 \rangle$ or $\mid 3 \rangle$  are accomplished by changing the
nuclear polarisation by $2M$ (so the final polarisation is either
$-M$ or $-M+2$). The transition to state $\mid 4 \rangle$  is made by
flipping a very large number $\sim N$ of nuclear spins, whilst still
changing the polarisation  by $2M$ only. This allows us to "fine tune"
a resonance with state $\mid 1 \rangle$, as described in the text. Only
a few polarisation groups are shown in the Figure; the insert shows
how these fit into the distribution $W( \epsilon )$ (shown for both initial
and final states).
}

\figure{{\bf Figure 5}  The same set of transitions as described in Fig.4,
but now for a grain in an external bias $\xi_H$, acting on $\vec{S}$.
The change in polarisation is $M_1+M_2$. The insert shows the initial
and final distributions $W(\epsilon \pm \xi_H)$, displaced from each other
by $2\xi_H$.
}

\figure{{\bf Figure 6} The time dependence of different contributions
$P_M(t)$ to the time correlation function $P(t)$ for an ensemble of
grains interacting only with nuclear spins, ignoring the effects
of nuclear spin diffusion. We assume that the parameter $\lambda =5$, i.e.,
roughly $5$ nuclei out of $N=1000$ are flipped each time $\vec{S}$ flips.
}

\figure{{\bf Figure 7} The time correlation function $P(t)$ for an
ensemble of grains, now plotted against $\ln t$; all contributions
$P_M(t)$ are included. This includes only nuclear spin bath-mediated
transitions. The figure describes either (a) $P(t)$ neglecting nuclear
spin diffusion, and having zero degeneracy blocking ($\mu =0$), or
(b), including both spin diffusion and finite degeneracy blocking.
}

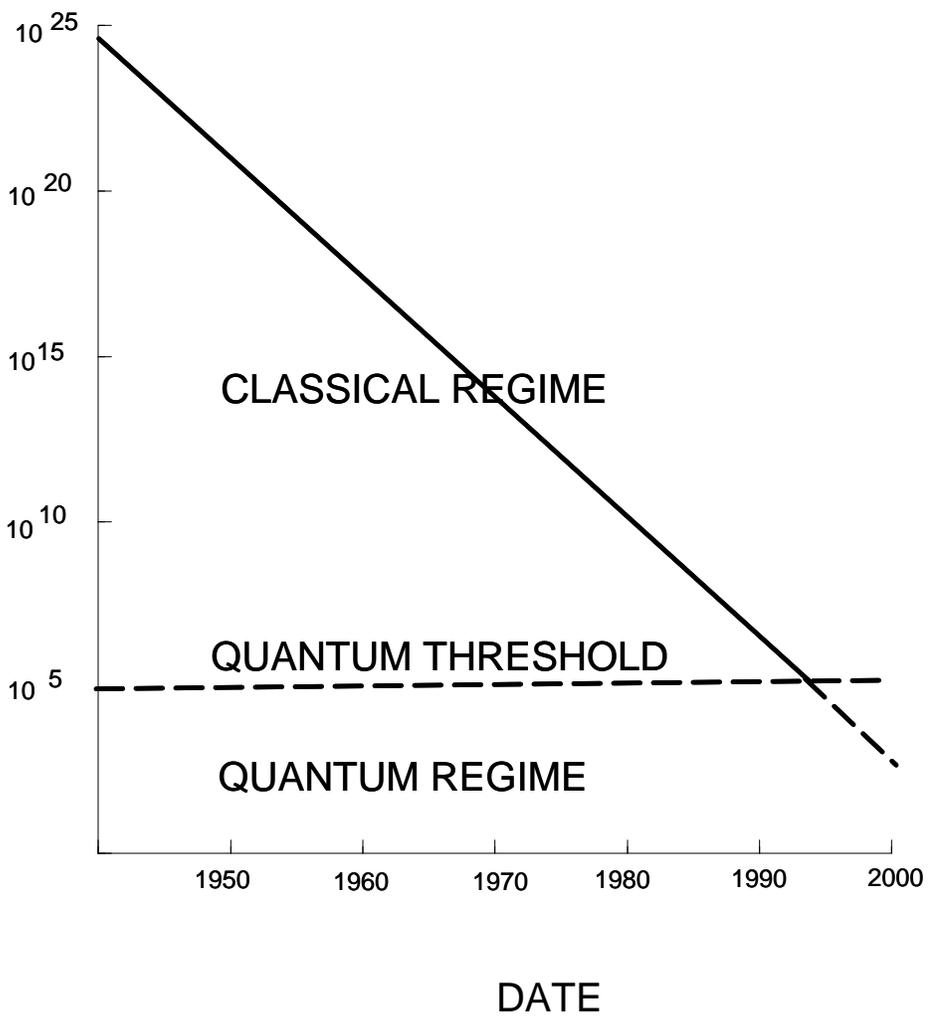
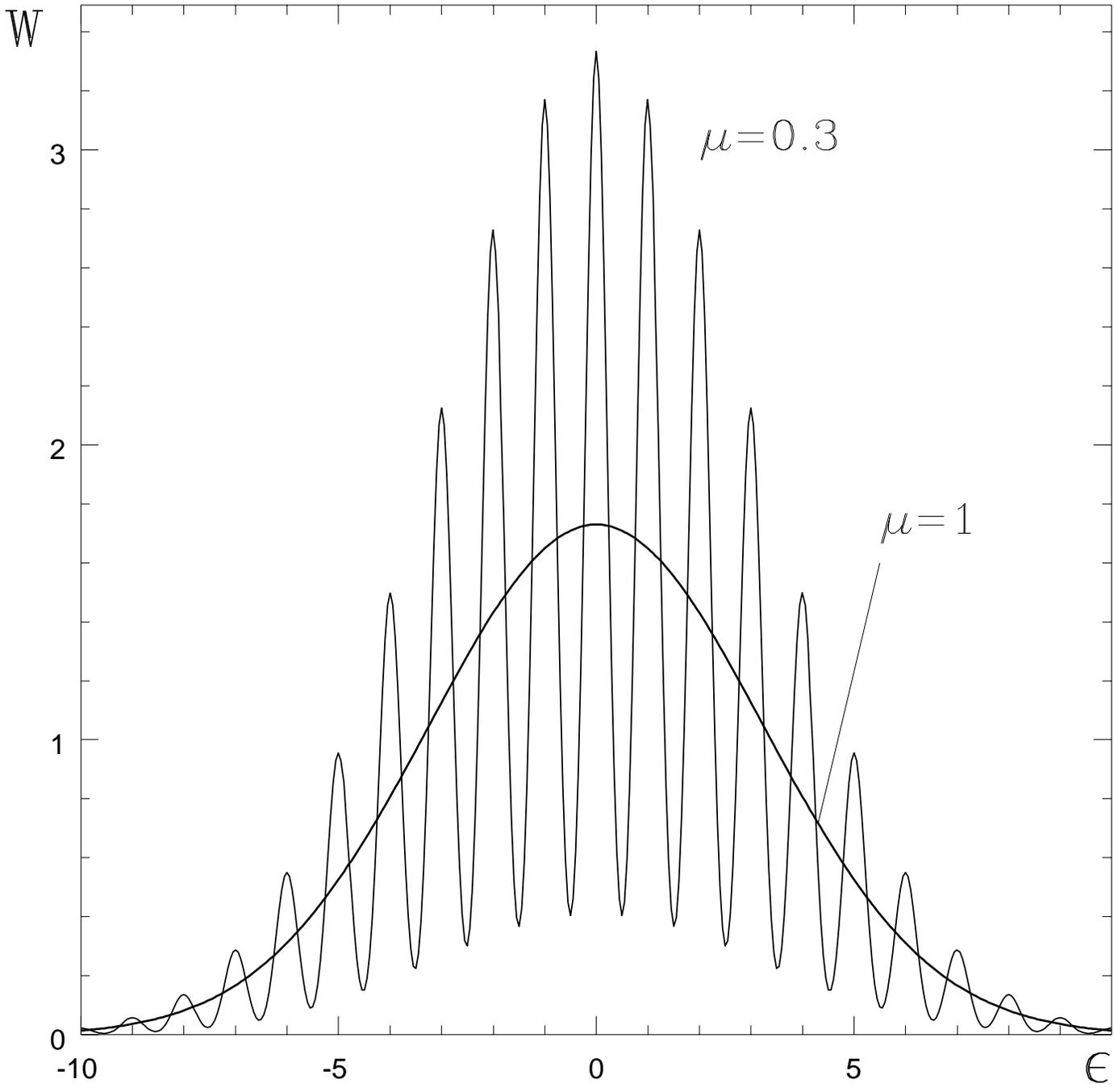
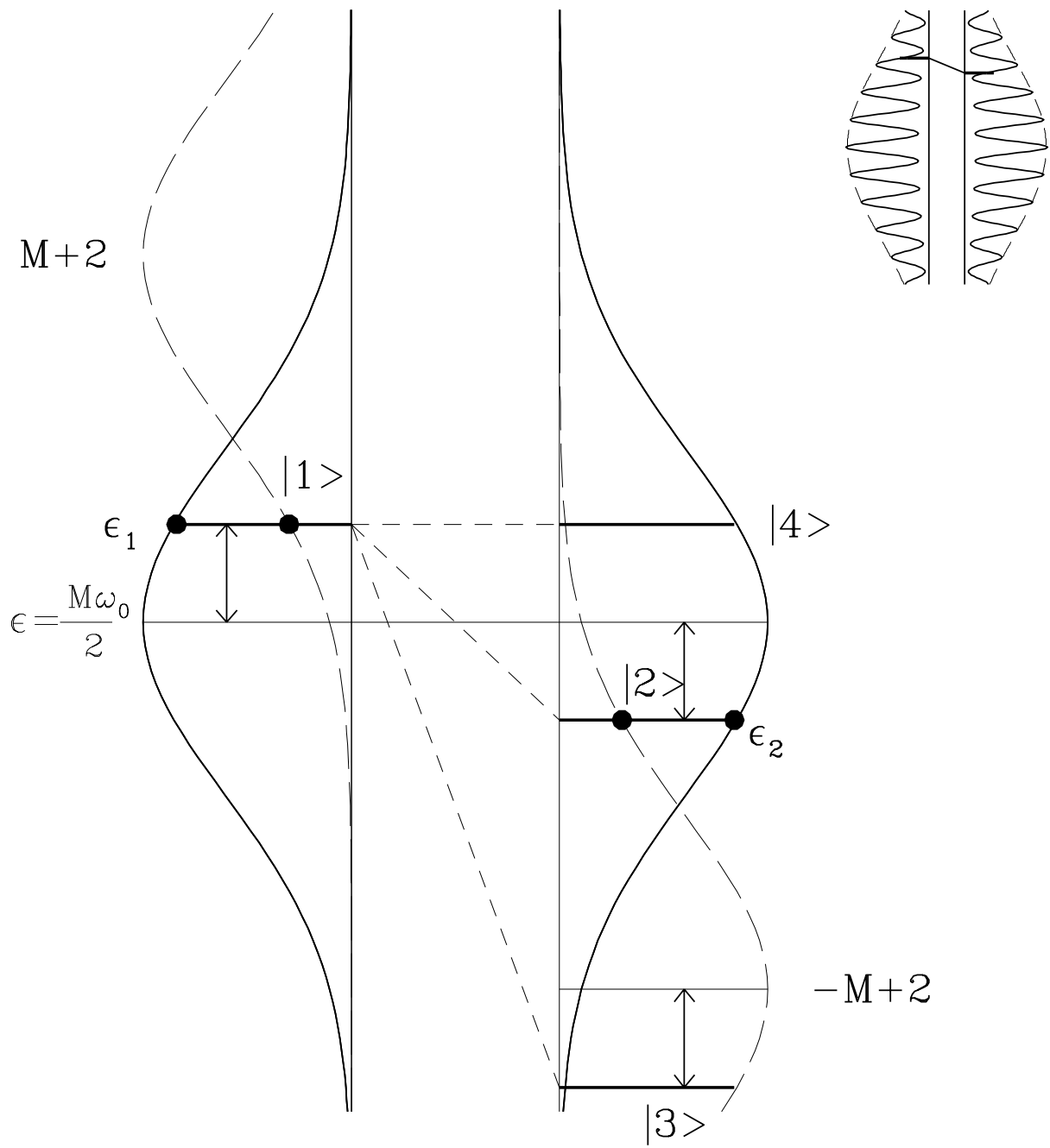
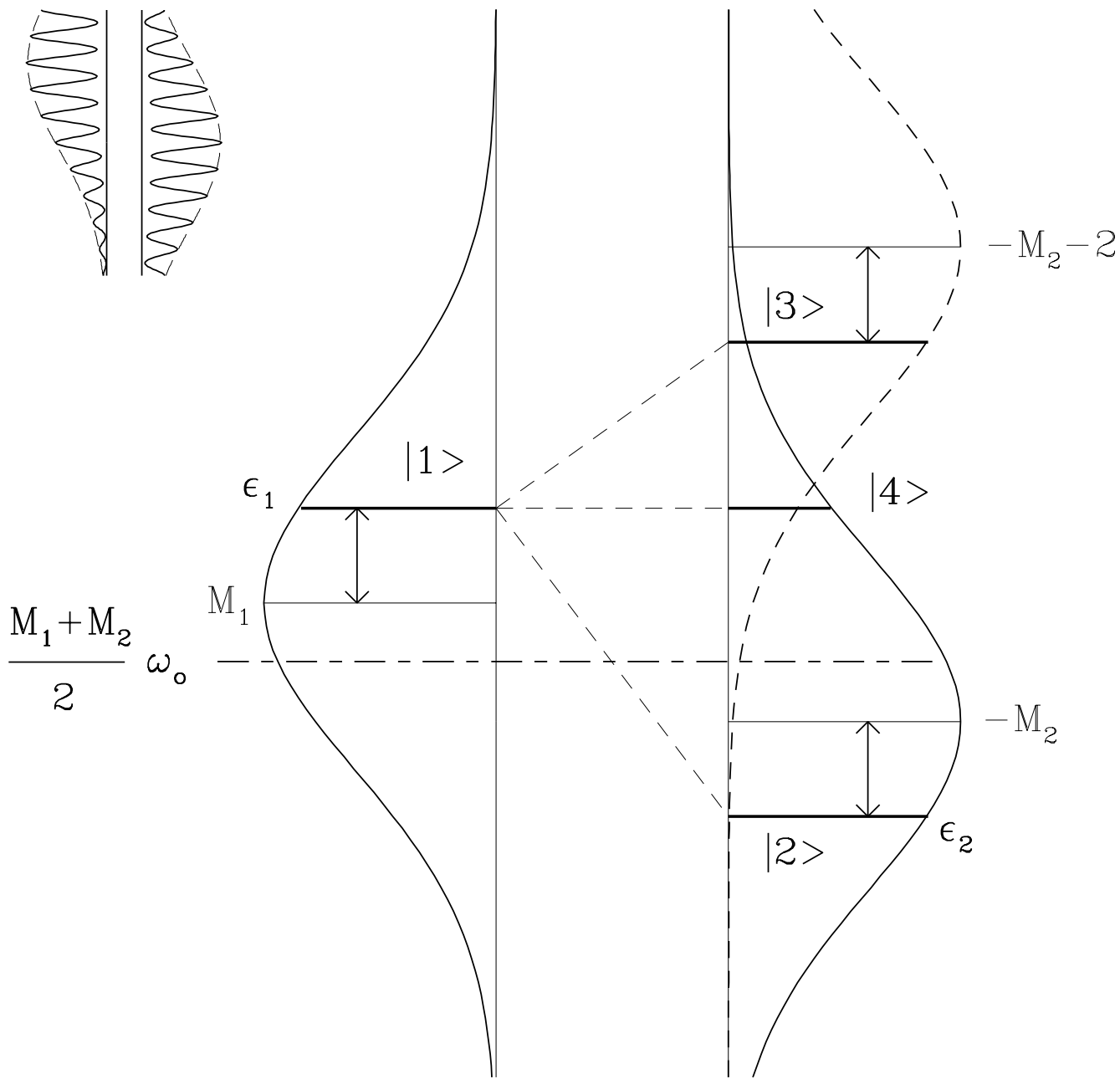
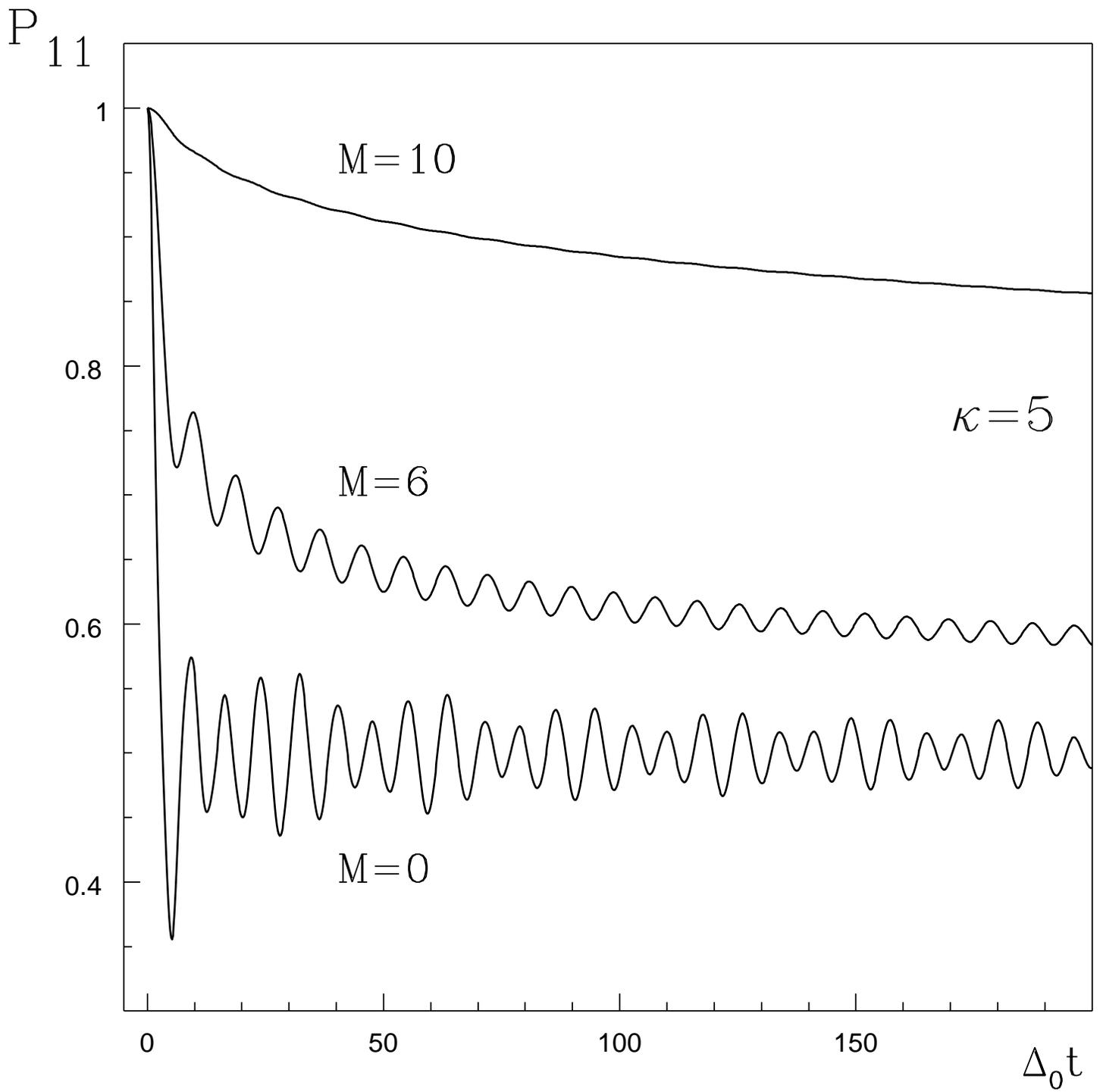
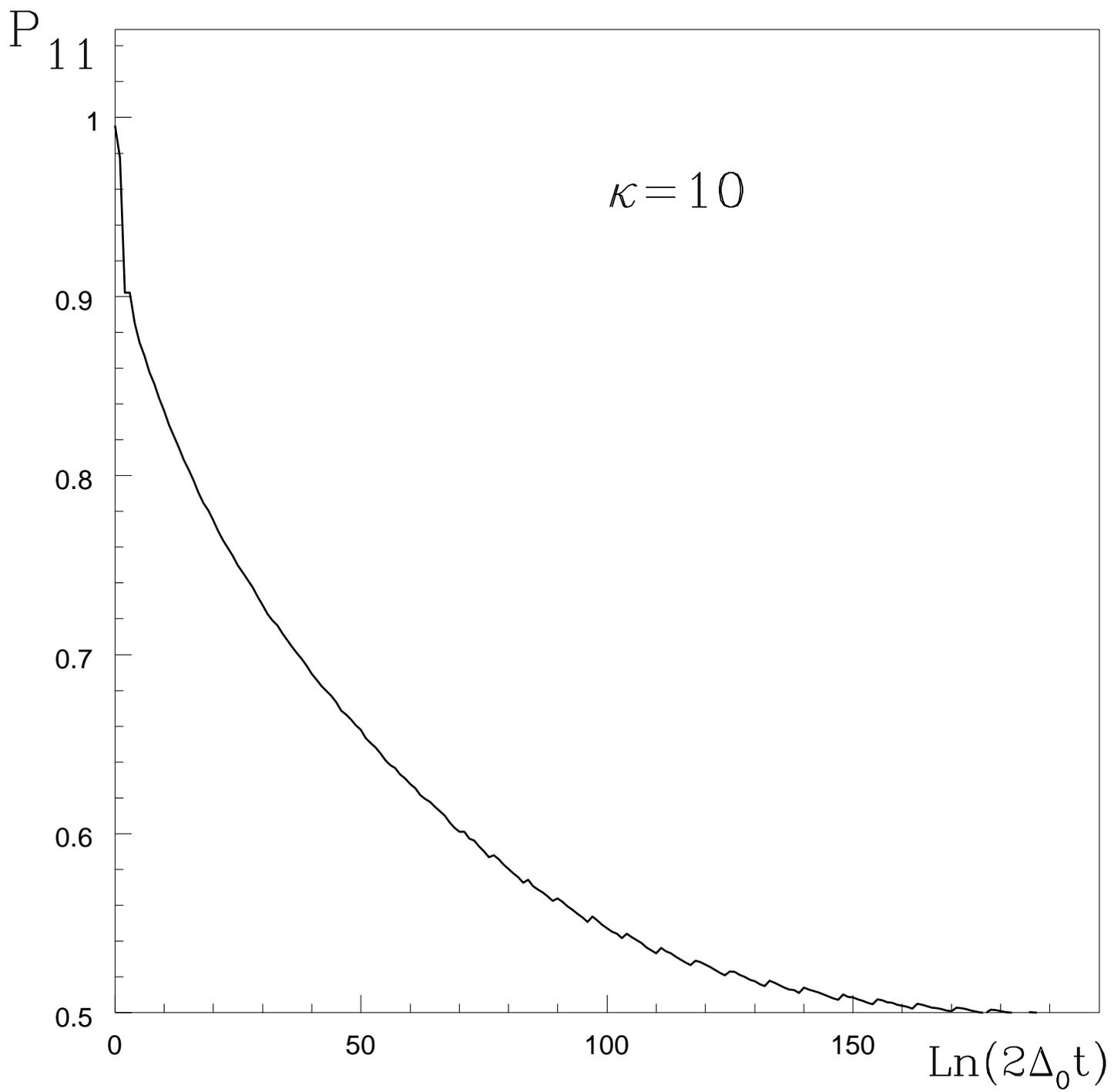
\end{document}